\documentclass[final,3p,times]{elsarticle}
\usepackage{amssymb}
\usepackage{amsmath}
\usepackage{algorithm}
\usepackage{algpseudocode}

\usepackage{multirow} 
\usepackage{booktabs}
\usepackage{float}
\usepackage{array}
\usepackage{wrapfig}
\usepackage{subcaption}  % 或者 \usepackage{subfigure}
\usepackage{enumitem}

\usepackage{listings}
\usepackage[usenames,dvipsnames]{xcolor}
\definecolor{mygreen}{rgb}{0,0.6,0}
\definecolor{mygray}{rgb}{0.5,0.5,0.5}
\definecolor{mymauve}{rgb}{0.58,0,0.82}
\journal{Journal of Computational Physics}

\begin{document}

\begin{frontmatter}



\title{An Improved High-order Adaptive Mesh Refinement Framework for Shock-turbulence Interaction Problems based on cell-centered finite difference schemes}

\author[inst1]{Yuqi Wang} %% Author name
\author[inst2]{Yadong Zeng}
\author[inst3]{Ralf Deiterding}
\author[inst4]{Jinhui Yang}
\author[inst1]{Jianhan Liang\corref{cor1}}
\cortext[cor1]{Correponding author: jhleon@vip.sina.com}

\affiliation[inst1]{organization={Advanced Propulsion Technology Laboratory, National University of Defense Technology},%Department and Organization
                    city={Changsha},
                    postcode={410073}, 
                    state={Hunan},
                    country={PR China}}

\affiliation[inst2]{organization={Department of Computer Science, University of Texas at Austin},%Department and Organization
                    city={Austin},
                    postcode={78712}, 
                    state={Texas},
                    country={USA}}
                    
\affiliation[inst3]{organization={AMROC CFD}, 
                addressline={Brookweg 167}, 
                city={Oldenburg}, 
                postcode={26127}, 
                country={Germany}}

\affiliation[inst4]{organization={College of Meteorology and Oceanography, National University of Defense Technology},%Department and Organization
                    city={Changsha},
                    postcode={410073}, 
                    state={Hunan},
                    country={PR China}}
% \affiliation[inst3]{organization={Computer, Computational, and Statistical Sciences Division,  Los Alamos National Laboratory}, 
%                 city={Los Alamos}, 
%                 postcode={87545}, 
%                 state={NM}, 
%                 country={USA}}

%% Abstract
\begin{abstract}
This work presents a high-order finite-difference adaptive mesh refinement (AMR) framework for robust simulation of shock-turbulence interaction problems. 
A staggered-grid arrangement, in which solution points are stored at cell centers instead of at the vertices, is presented to address the boundary conservation issues encountered in previous studies. 
The key ingredient in the AMR framework, i.e., the high-order nonlinear interpolation method applied in the prolongation step together with the determination of fine-grid boundary conditions, are re-derived for staggered grids following the procedures in prior work~\cite{sebastianMultidomainWENOFinite2003} and are thus used here.
Meanwhile, a high-order restriction method is developed in the present study as the coarse and fine grid solutions are non-collocated in this configuration.
To avoid non-conservative interpolation at discontinuous cells that could incur instabilities, a hybrid interpolation strategy is proposed in this work for the first time, where the non-conservative WENO interpolation is applied in smooth regions whereas the second-order conservative interpolation is applied at shocks.
This significantly mitigates the numerical instabilities introduced by non-conservative interpolation and pointwise replacement.
The two interpolation approaches are seamlessly coupled through a troubled-cell detector achieved by a scale-irrelevant Riemann solver in a robust way.
The present work is developed on a publicly available block-structured adaptive mesh refinement framework AMReX~\cite{zhangAMReXFrameworkBlockstructured2019}. 
The canonical tests demonstrate that the proposed method is capable of accurately resolving a wide range of complex shock–turbulence interaction problems that have been proven intricate for existing approaches.
\end{abstract}

\begin{graphicalabstract}
\end{graphicalabstract}

\begin{highlights}

\item A cell-centered adaptive mesh refinement (AMR) framework is implemented that simultaneously maintains high-order accuracy and strict flux conservation at coarse-fine interfaces, resolving the fundamental trade-off between accuracy and conservation that plagues traditional vertex-centered AMR methods.

\item Optimally fifth-order central WENO interpolation schemes are introduced additionally for the restriction step in AMR, ensuring order consistency while maintaining essentially non-oscillatory properties.

\item A Hybrid WENO interpolation strategy is developed that combines troubled-cell detection with adaptive interpolation schemes: high-order WENO interpolation in smooth regions and robust conservative treatment near discontinuities, ensuring both accuracy and stability across complex flow features.

\end{highlights}

%% Keywords

\end{frontmatter}



\section{Introduction}
\label{sec1}
Simulations of high-Reynolds turbulent flows pose formidable numerical challenges due to the presence of multiscale structures, such as the shock-turbulence interaction, which could span several orders of magnitude in space and time. 
Resolving the dynamic structures with a static, uniform mesh demands an enormous computational cost, which exceeds the computation capacity of current computers.
In contrast to it, adaptive mesh refinement (AMR) offers substantially higher efficiency by concentrating the grid points on localized regions with larger truncation errors, which makes it a powerful tool for high-resolution simulations.

The adaptive mesh refinement (AMR) method was originally developed for hyperbolic conservation laws~\cite{bergerAdaptiveMeshRefinement1984,bergerLocalAdaptiveMesh1989}. 
Since then, it has been extensively extended and applied to a wide range of problems, including fluid–structure interaction, aerodynamics, magnetohydrodynamics, and combustion. 
Early AMR implementations were primarily designed to achieve second-order accuracy, which is proved highly successful in shock-dominated laminar flows~\cite{deiterdingParallelAdaptiveSimulation,sunDetonationFoamOpensourceSolver2023,wangNumericalStudyRapid2022}. 
In such cases, second-order AMR schemes offered a favorable compromise between accuracy and computational costs.
However, when applied to broadband turbulence simulations, the efficiency of low-order AMR method deteriorates significantly. 
Jameson~\cite{jamesonAMRVsHigh2003} has found out that to resolve turbulence, it is often more efficient to employ high-order schemes on uniform grids rather than using second-order AMR grids, and that only when AMR methods achieve fourth-order accuracy or higher can they regain competitiveness for such problems.

High-order AMR methods were developed earlier particularly based on the finite-volume (FV) framework. 
To preserve the global accuracy, apart from the spatial and temporal discretizations, interpolation between the coarse and fine grids in AMR must also retain high-order.
McCorquodale and Colella~\cite{mccorquodaleHighorderFinitevolumeMethod2011} pioneered a high-order FV-AMR algorithm on hyperbolic conservation laws with a fourth-order, conservative least-squares approach for the coarse-fine interpolation, which addresses the accuracy degeneration in AMR  in a conservation form. 
% However, they employed an artificial viscosity technique at shock discontinuities, which tends to introduce excessive numerical dissipation.
Later on, Zhang et al generalized this framework to elliptic and parabolic problems, enabling the algorithms applicable to more complex systems.
By including the chemistry, Emmett et al~\cite{emmettFourthOrderAdaptiveMesh2019} further presented a fourth-order AMR algorithm for solving reactive flows, with the usage of a high-order conservative quartic polynomial interpolation scheme for prolongation.
The restriction procedure, i.e., synchronizing the fine grids onto the coarse grids, was implemented by volume-weighted averaging.
Other investigations also focus on, for example curvilinear grids~\cite{chesshireSchemeConservativeInterpolation1994,mignoneHighorderConservativeReconstruction2014a} and particular prolongation schemes~\cite{buchmullerFiniteVolumeWENO2016a,reevesApplicationGaussianProcess2022}, just to name a few.
However, the flux reconstruction process in FV introduces substantial computational costs especially in higher dimensions, thereby not suitable for large-scale turbulence modeling. 
Meanwhile, the use of multi-dimensional least-squares projection technique involves solving a linear equation system at runtime, which further exacerbates the situation.
While some modified algorithms were presented to alleviate this issue, for example, by reducing the coarse–fine interpolation to matrix–vector products~\cite{zhangHighorderMultidimensionalConservative2011} or using a Gaussian Processing Modeling approach~\cite{reevesApplicationGaussianProcess2022}, the overall efficiency of such approaches remains fundamentally constrained by the finite-volume framework itself.
In addition to that, the interpolation techniques are algorithmically intricate, making them difficult to implement and port efficiently across different platforms.

In contrast, it is well known that the finite difference (FD) discretization methods have superior computational efficiency and compactness, particularly for high-dimensional turbulence simulations on structured meshes. 
As noted by Shu~\cite{shuHighorderFiniteDifference2003a}, the weighted essentially non-oscillatory schemes, denoted as WENO, in the FD formulation can save roughly 4× in 2D and 9× in 3D computational costs than their FV counterparts.
The original FD-WENO scheme was developed by Jiang and Shu~\cite{jiangEfficientImplementationWeighted1996}, which is also known as the WENO-JS scheme.
An obvious drawback of this scheme is its loss of accuracy at critical points, which may lead to excessive damping of small-scale features.
% Later on, Henrick et al~\cite{henrickMappedWeightedEssentially2005} noticed that the WENO-JS scheme may loss accuracy at critical points and thus can deteriorate to evaluate small scale structures in turbulent flows.
% To address this issue, they developed the WENO-M scheme but the computational cost is increased significantly.
% Subsequently, Borges et al~\cite{borgesImprovedWeightedEssentially2008} presented the WENO-Z scheme, which can preserve the accuracy on critical points while significantly reduces the workload compared to the WENO-M scheme.
% Other schemes that were recently developed, for example the WENO-CU scheme~\cite{huAdaptiveCentralupwindWeighted2010} developed for improve the dissipation at smooth regions and the 
Later on, a variety of improved WENO schemes were developed, e.g., for achieving desired order at critical points~\cite{henrickSimulationsPulsatingOnedimensional2006, borgesImprovedWeightedEssentially2008}, for improving resolution in smooth regions~\cite{huAdaptiveCentralupwindWeighted2010,zhuNewFifthOrder2016} and for suppressing spurious oscillations near discontinuities~\cite{balsaraMonotonicityPreservingWeighted2000}, and so on.
% In addition to that, Fu et al in~\cite{fuFamilyHighorderTargeted2016} further improved the scheme's robustness and dissipation properties by proposing a family of high-order targeted ENO schemes (TENO) with an additional ENO-like stencil selection concept.
% Balsara et al~\cite{balsaraEfficientClassWENO2016} also presented an efficient class of adaptive-order WENO (WENO-AO) schemes by expressing the stencils in compact Legendre polynomials and non-linearly hybridizing a very high order stencil with an third-order CWENO scheme.
% Recently, Zhu and Shu~\cite{zhuNewTypeMultiresolution2018} proposed the WENO-MR scheme with the usage of a hierarchy of unequal-size central stencils, substantially avoiding accuracy loss at critical points and ensuring the numerical robustness.
Recent efforts on the improvement of WENO schemes include the TENO~\cite{fuFamilyHighorderTargeted2016} scheme, the WENO-AO~\cite{balsaraEfficientClassWENO2016} scheme and the WENO-MP~\cite{zhuNewTypeMultiresolution2018} scheme, just to name a few.
Notably, although WENO schemes have demonstrated remarkable effectiveness for turbulence simulations on uniform grids~\cite{borgesImprovedWeightedEssentially2008,zhuNewTypeMultiresolution2018}, their integration with AMR poses additional challenges, which mainly lie in the exchange of information between coarse and fine grids while ensuring stability, accuracy, and conservation.

In the context of AMR, its grids can be regarded essentially as a hierarchy of non-conforming grid patches, with additionally the finer grids populated in sequence from its next coarse grids.
Sebastian and Shu~\cite{sebastianMultidomainWENOFinite2003} first proposed using a spatial interpolation procedure at the subdomain interfaces within the multidomain high-order FD-WENO method.
Their numerical experiments demonstrated that the Lagrange interpolation was more efficient and accurate enough for the treatment of subdomain interfaces, as compared to the nonlinear WENO interpolation scheme, with the spurious oscillations arisen from the interpolation procedure sufficiently suppressed during spatial discretization, i.e., by the characteristic decomposition and nonlinear stencil assembly during the WENO reconstruction.
Nevertheless, both interpolation methods exhibited considerable conservation errors, especially when the shock discontinuities crossed the subdomain interfaces, which may trigger numerical instabilities and degrade the robustness of the overall algorithms. 
Li and Hyman~\cite{liAdaptiveMeshRefinement} subsequently proposed using spatially second-order conservative interpolation in AMR grids to retain the numerical conservation, which nevertheless remains a compromise as it substantially damps the small-scale structures in smooth regions. 

Recognizing that any high-order interpolation in the finite-difference framework is inherently non-conservative, Shen et al.~\cite{shenAdaptiveMeshRefinement2011a} adopted the WENO interpolation approach of~\cite{sebastianMultidomainWENOFinite2003} for AMR computations combined with a high-order FD-WENO scheme, thereby trading strict conservation for the ability to retain optimal order in smooth regions.
They assumed the discontinuities always tagged and kept refined at each time step, so that the resulting conservation errors remain acceptable.
In practice, however, this assumption requires discontinuities be strictly confined within the finest grids at all times—a condition that is difficult to guarantee in practical usage, particularly when a new refinement region is created or when the shock crosses the coarse–fine interfaces.
If this condition is violated, conservation errors may accumulate at interfaces and potentially lead to numerical instabilities.
This interpolation procedure was also applied for reactive flows in~\cite{chenFifthorderFiniteDifference2016,raiConservativeHighorderaccurateFinitedifference}, while the conservation issue remained still unresolved.

In summary, a key bottleneck in developing high-order AMR schemes lies in designing interpolation strategies that consistently preserve accuracy while maintaining conservation, particularly in regions with the presence of discontinuities. 
Finite-volume approaches address this challenge through constrained least-squares reconstructions~\cite{zhangHighorderMultidimensionalConservative2011,zhangFourthOrderAccurateFiniteVolume2012,buchmullerFiniteVolumeWENO2016a,emmettFourthOrderAdaptiveMesh2019}, but at the expense of considerable algorithmic complexity and computational cost in higher dimensions.
In contrast, several high-order FD-WENO based AMR methods~\cite{shenAdaptiveMeshRefinement2011a,zieglerAdaptiveHighorderHybrid2011,houimLowdissipationTimeaccurateMethod2011} can retain optimal order of accuracy in smooth regions, yet significant concerns remain regarding their accuracy and stability when applied to shock-dominated flows.
Moreover, in previous FD-WENO AMR frameworks, the solution points on both coarse and fine grids are defined at the cell vertices~\cite{shenAdaptiveMeshRefinement2011,wangParallelAdaptiveMesh2015a,liAdaptiveMeshRefinement}, which is aligned with the original WENO-JS scheme~\cite{jiangEfficientImplementationWeighted1996}.
Since the coarse grid points are collocated with the finer ones generated during the mesh refinement, the updated finer grid values are typically copied directly to the corresponding coarse grid points. 
Although efficient, this direct point-value replacement would nevertheless violate the mass conservation on the coarse grid, and thus is unfavorable for robust computations with strong shocks.

% Non-conservative interpolation may lead to noticeable conservation errors, particularly in shock-dominated flows where a physical shock speed is crucial.
% By collocating the coarse and fine patches at the solution points, their method also struggled to simultaneously preserve the accuracy and conservation at the interface.
% In cell-centered FD schemes, such as the one considered in this work, fine-to-coarse synchronization (i.e., restriction) is additionally considered such that the conservation is straightforward to implement by a flux correction step without compromising the formal accuracy. 

In this work, we develop an improved AMR-WENO algorithm on top of the AMReX~\cite{zhangAMReXFrameworkBlockstructured2019}.
Our interpolation strategy is primarily based on a hybrid scheme, in which a high-order centered interpolation is applied in smooth regions, while the scheme adaptively switches to second-order conservative interpolation near shock discontinuities.
The explicit switching is guided by a troubled-cell detector based on an approximate Riemann solver. 
Overall, this hybrid approach balances the accuracy and the conservation, leading to enhanced robustness. 
Moreover, differing from the typical vertex-based finite-difference schemes, we adapt the classic WENO formulation into a staggered arrangement, in which the solution points are located at cell centers and flux points are at cell interfaces, like the weighted nonlinear compact (WCNS)  scheme~\cite{dengDevelopingHighOrderWeighted2000a,chenNonlinearWeightsShock2023} and the spectral difference scheme~\cite{koprivaConservativeStaggeredGridChebyshev1996}. 
Although this arrangement appears similar to the finite-volume methods, the key distinction is that the former remains storing pointwise values while the latter stores cell averages. 
However, this setup enables strict conservation at the coarse–fine interfaces, which was difficult to retain for previous vertex-based AMR implementations. 
In return, since the coarse solution points are not collocated with the finer solution points anymore in this configuration, we also develop an oscillation-free interpolation strategy necessarily for the restriction procedure with optimally fifth-order of accuracy in smooth regions, resorting also to the aforementioned troubled-cell detector technique.
For practical usage, we derive and extend the entire algorithms to handle both 2- and 4-refinement ratios, thereby enhancing the flexibility of our high-order AMR implementation. 
Compared to previous studies~\cite{zieglerAdaptiveHighorderHybrid2011,houimLowdissipationTimeaccurateMethod2011,shenAdaptiveMeshRefinement2011}, the present scheme offers improved accuracy and stability for turbulence-resolving simulations involving strong shocks, which is shown in Section~\ref{sec6}.

% The algorithms in this paper is developed on top of  AMReX~\cite{zhangAMReXFrameworkBlockstructured2019}, a high-performance block-structured adaptive mesh framework that has been extensively used for the development of application codes, which we have used for developing our prior framework for reactive flows in second-order accuracy~\cite{wangAdaptiveSolverAccurate2025}.

The paper is organized as follows: Section 2 introduces the governing equations for three-dimensional Navier-Stokes equations; Section 3 details the spatial and temporal discretization schemes for the governing system; Section 4 presents the key ingredients of our AMR-WENO algorithms, including high-order spatiotemporal interpolation methods and flux correction algorithms; 
Section 5 presents the key ingredient of hybrid interpolation strategy;
% Section 5 explores GPU optimization strategies for our algorithm; 
The numerical tests are investigated and analyzed in Section 6; and finally, Section 7 summarizes the conclusion and current limitations.

\section{Governing Equations}
\label{sec2}
This work focuses on solving the Navier-Stokes gas dynamics equations, which can be written in conservative form as
\begin{gather}
\label{eq:governing}
    \frac{\partial \boldsymbol{u}}{\partial t} + \frac{\partial f}{\partial x} + \frac{\partial g}{\partial y} + \frac{\partial h}{\partial z} = \frac{\partial f^v}{\partial x} + \frac{\partial g^v}{\partial y} + \frac{\partial h^v}{\partial z} \;, 
\end{gather}
\begin{equation}
\label{eq:state vector}
    \boldsymbol{u} = \left[\rho, \rho u, \rho v, \rho w, \rho E\right]^{T},
\end{equation}
where the pressure is given by 
\begin{equation}
    p = (\gamma-1)\rho (E - \frac{u^2+v^2+w^2}{2})
\end{equation}
with $\gamma=1.4$ for a perfect gas. 
The convective fluxes are defined as
\begin{gather}
\label{eq:convective_flux}
    f = \begin{pmatrix}
        \rho u \\
        \rho u^2 + p \\
        \rho u v \\
        \rho u w \\
        (\rho E + p)u \\
    \end{pmatrix}, \quad
    g = \begin{pmatrix}
        \rho v \\
        \rho u v \\
        \rho v^2 + p \\
        \rho v w \\
        (\rho E + p)v \\
    \end{pmatrix}, \quad 
    h = \begin{pmatrix}
        \rho w \\
        \rho u w \\
        \rho v w \\
        \rho w^2 + p \\
        (\rho E + p)w \\
    \end{pmatrix}, 
\end{gather}
The viscous fluxes are given by
\begin{equation}
\label{eq:viscous_flux}
\begin{aligned}
    f^{v}=
        \begin{pmatrix}
        0 \\
        \tau_{xx} \\
        \tau_{xy} \\
        \tau_{xz} \\
        u\tau_{xx} + v\tau_{xy} + w\tau_{xz} - q_x\\
        \end{pmatrix}, \quad
    g^{v}=
        \begin{pmatrix}
        0 \\
        \tau_{yx} \\
        \tau_{yy} \\
        \tau_{yz} \\
        u\tau_{yx} + v\tau_{yy} + w\tau_{yz} - q_y\\
        \end{pmatrix}, \quad 
    h^{v}=
        \begin{pmatrix}
        0 \\
        \tau_{zx} \\
        \tau_{zy} \\
        \tau_{zz} \\
        u\tau_{zx} + v\tau_{zy} + w\tau_{zz} - q_z\\
        \end{pmatrix}.
\end{aligned}
\end{equation}
The viscous stresses tensor $\tau$ is given by
\begin{equation}
\label{eq:viscous_stress}
    \tau=-\frac{2}{3} \mu(\nabla \cdot \boldsymbol{v}) \mathbf{I}+\mu\left[\nabla \boldsymbol{v}+(\nabla \boldsymbol{v})^{T}\right],
\end{equation}
where the shear viscosity $\mu$ is obtained from the Sutherland formula.
The heat conduction vector $q$ is given by Fourier's law
\begin{equation}
\label{eq:heat_conduction}
    q=-\lambda \nabla T
\end{equation}
where $\lambda$ is the thermal conductivity, which is determined by the Prandtl number.
The above system is closed by the ideal gas equation of state
\begin{equation}
    p = \rho R T,
\end{equation}
where $R$ denotes the specific gas constant.

% \begin{equation*}
% \mu=\mu_{0} \frac{T_{0}+C}{T+C}\left(\frac{T}{T_{0}}\right)^{\frac{3}{2}}, \quad \lambda=\frac{c_{p} \mu}{\operatorname{Pr}}=\frac{\gamma R}{(\gamma-1)} \frac{\mu}{\operatorname{Pr}}
% \end{equation*}
% Here, the Prandtl number Pr is set as respectively 0.72 by default, corresponding to the value of the air at 1 atm and between 200 K-1000 K, $R$ is 287 J/(kg·K) and the Sutherland constants $C$ = 110.5 K, $T_{0}$ = 273.1 K and $\mu_{0} = 1.716\times10^{-5}$  Pa$\cdot$s, respectively.

\section{Single-level numerical method}
\label{sec3}
WENO schemes have become a standard tool for compressible turbulence simulations, combining low numerical dissipation in smooth regions with robust shock-capturing capability. Although many improved variants have been proposed, we employ the original Jiang-Shu-type nonlinear weights~\cite{jiangEfficientImplementationWeighted1996a} in both the spatial discretization and AMR interpolation methods in this work, as the focus of this study is on developing a general methodology for establishing high-order AMR–WENO algorithms rather than on improving the WENO schemes themselves.

\subsection{Spatial discretization}
\label{sec3.1}
The governing Eqs \ref{eq:governing} can be discretized line by line in the Cartesian coordinates $(x,y,z)$. 
Therefore, in this section we only consider the one-dimensional discretization method for simplification.
The computational domain $[a,b]$ is divided into $N$ cells by the flux points
$
x_{i+1/2} = a + i \Delta x, \ i=0,1,\ldots,N,
$
with $\Delta x = (b-a)/N$ the grid spacing. 
In this work, the solution points are placed at the cell centers,
$
x_j = a + \left(i-\tfrac{1}{2}\right)\Delta x, \ i=1,2,\ldots,N.
$
In this case, the domain boundaries are flux points rather than solution points. 
This staggered-grid setup was used previously in Weighted Compact Nonlinear Schemes and spectral difference schemes, primarily for avoiding corner issues in multi-domain problems.
The setup is further leveraged in a novel way by us to ensure local conservation at coarse–fine interfaces, as detailed in Section~\ref{sec4.1}.

For spatial discretization, we adopt the classical fifth-order WENO-JS reconstruction. 
For each flux point at $x_{i+1/2}$, on a biased stencil $\{x_{i-2},x_{i-1},x_i,x_{i+1},x_{i+2}\}$, three candidate substencils
$
S_0=\{f_{i-2},f_{i-1},f_i\}, \ 
S_1=\{f_{i-1},f_i,f_{i+1}\}, \
S_2=\{f_i,f_{i+1},f_{i+2}\}
$
are used to approximate the flux at the interface $f_{i+1/2}$. 
The smoothness indicators are given by
\begin{align}
\beta_0 &= \tfrac{13}{12}(f_{i-2}-2f_{i-1}+f_i)^2 + \tfrac{1}{4}(f_{i-2}-4f_{i-1}+3f_i)^2, \\
\beta_1 &= \tfrac{13}{12}(f_{i-1}-2f_i+f_{i+1})^2 + \tfrac{1}{4}(f_{i-1}-f_{i+1})^2, \\
\beta_2 &= \tfrac{13}{12}(f_i-2f_{i+1}+f_{i+2})^2 + \tfrac{1}{4}(3f_i-4f_{i+1}+f_{i+2})^2.
\end{align}
The nonlinear weights follow
\begin{equation}
\omega_k = \frac{\alpha_k}{\sum_{j=0}^2 \alpha_j}, \quad
\alpha_k = \frac{d_k}{(\beta_k+\epsilon)^2}, \quad
(d_0,d_1,d_2)=\left(1/10, 6/10, 3/10\right),
\end{equation}
leading to the reconstructed flux
\begin{equation}
f_{i+1/2} = \sum_{k=0}^2 \omega_k q^{(k)}_{i+1/2},
\end{equation}
where $q^{(k)}_{i+1/2}$ are the $k$-th candidate polynomials. 
This achieves fifth-order accuracy in smooth regions and suppresses oscillations near discontinuities.

To further alleviate numerical oscillations, the reconstruction is performed in characteristic space using Roe-averaged Jacobians. 
The fluxes are decomposed into characteristic components
$
\tilde{f}_k = L_{i+1/2} f_k, \quad \tilde{u}_k = L_{i+1/2} u_k,
$
where $L_{i+1/2}$ denotes the left eigenvectors of the Jacobian $A_{i+1/2}=\partial f/\partial u$. 
A local Lax–Friedrichs splitting is applied,
$
\tilde{f}^\pm = \tfrac{1}{2}(\tilde{f} \pm \alpha \tilde{u}),
$
with $\alpha$ the upper-bound wave speed.
The WENO reconstruction is carried out separately for $\tilde{f}^+$ and $\tilde{f}^-$, followed by projection back to physical space:
$
f_{i+1/2}^\pm = R_{i+1/2}\tilde{f}_{i+1/2}^\pm, 
$
where $R_{i+1/2}$ is the matrix of right eigenvectors. 
The final numerical flux is obtained by
$
\hat{f}_{i+1/2} = f_{i+1/2}^+ + f_{i+1/2}^-.
$

\subsection{Temporal scheme}
\label{sec3.2}
To evolve the solution in time, we employ the third-order strong stability preserving Runge-Kutta (SSP-RK3) method~\cite{gottliebTotalVariationDiminishing1998}
\begin{equation}
\begin{aligned}
\boldsymbol{u}^{(1)} &= \boldsymbol{u}^{n} + \Delta t \, \boldsymbol{k}(\boldsymbol{u}^{n},t^n), \\
\boldsymbol{u}^{(2)} &= \frac{3}{4} \boldsymbol{u}^{n} + \frac{1}{4} \boldsymbol{u}^{(1)} + \frac{1}{4} \Delta t \, \boldsymbol{k}(\boldsymbol{u}^{(1)},t^n+\Delta t), \\
\boldsymbol{u}^{n+1} &= \frac{1}{3} \boldsymbol{u}^{n} + \frac{2}{3} \boldsymbol{u}^{(2)} + \frac{2}{3} \Delta t \, \boldsymbol{k}(\boldsymbol{u}^{(2)},t^n+\frac{1}{2}\Delta t),
\end{aligned}
\end{equation}
where $\boldsymbol{k}(\boldsymbol{u})$ denotes the spatial discretization operator.

To ensure numerical stability, the timestep $\Delta t$ is restricted according to the CFL condition derived from multi-dimensional Navier-Stokes analysis, following~\cite{zieglerAdaptiveHighorderHybrid2011}.

\section{Block-structured Adaptive Mesh Refinement (SAMR)}
\label{sec4}

% \begin{figure}
%     \centering
%     \includegraphics[width=0.5\linewidth]{figs/AMR_algorithms.pdf}
%     \caption{Caption}
%     \label{fig:placeholder}
% \end{figure}

% \begin{figure}
%     \centering
%     \includegraphics[width=0.4\linewidth]{figs/AMR_subcycling.pdf}
%     \caption{Caption}
%     \label{fig:placeholder}
% \end{figure}

\begin{figure}
    \centering
    \includegraphics[width=0.99\linewidth]{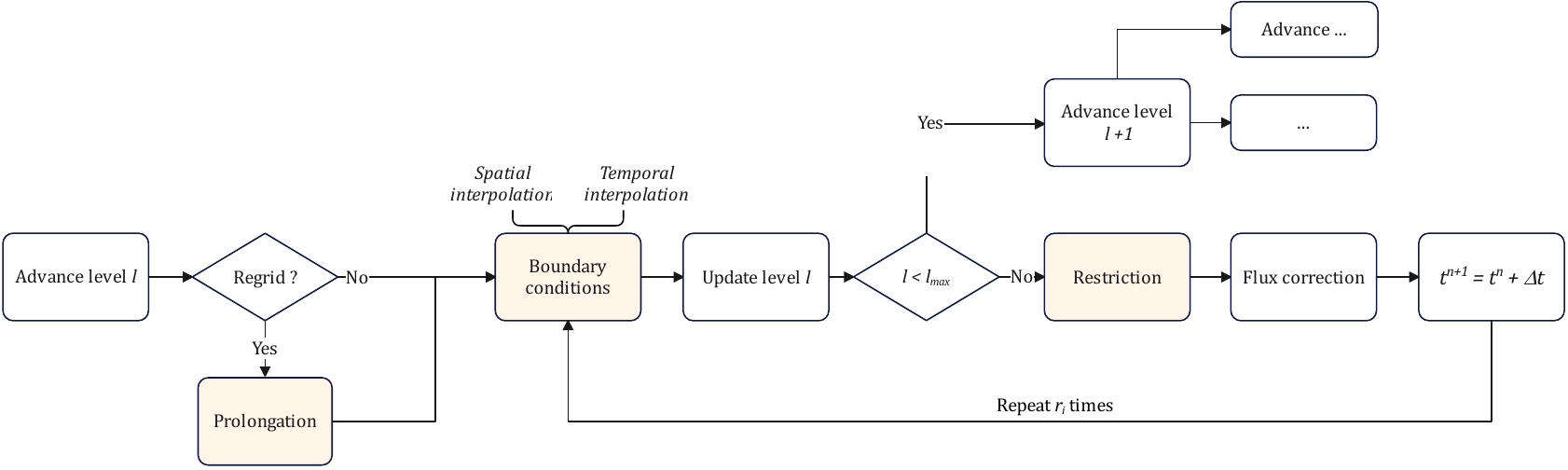}
    \caption{Algorithmic flowchart of the subcycling AMR framework. The highlighted rounded rectangles indicate the procedures where a high-order interpolation schemes are required.}
    \label{fig:flow_chart_AMR}
\end{figure}

Although the fundamental ingredients for block-structured AMR have been well established in the literature~\cite{bergerAdaptiveMeshRefinement1984,deiterdingAdaptiveMultiresolutionAdaptive2009}, an improved high-order framework for solving multiphysics gas-dynamics problems is presented here.
Figure~\ref{fig:flow_chart_AMR} illustrates the core workflow of a traditional finite-volume AMR algorithm. 
In this work, we adapt it to a high-order finite-difference AMR-WENO framework on staggered grids.

The open-source AMReX library~\cite{zhangAMReXFrameworkBlockstructured2019} is employed as the underlying infrastructure, providing standard AMR components such as grid nesting, dynamic regridding, ghost-cell filling, and basic coarse–fine synchronization.
However, what we mainly focus in the present work are the necessary procedures required to achieve high-order accuracy, which are highlighted in Figure~\ref{fig:flow_chart_AMR}.
Note, that unlike previous finite-difference AMR frameworks~\cite{shenAdaptiveMeshRefinement2011,wangParallelAdaptiveMesh2015a}, our general methodology also incorporates a flux correction step, thereby ensuring the conservation at coarse-fine interfaces.

% \begin{figure}[htpb]
%     \centering
%     \includegraphics[width=0.9\linewidth]{figs/AMR_methodology.pdf}
%     \caption{Coarse-fine synchronization with information transferred from (a) the coarse grid $G^{l-1}$ to the fine grid $G^l$ (on both valid and ghost cells) and (b) the fine grid $G^l$ to the coarse grid $G^{l-1}$ (on valid cells only).}
%     \label{fig:AMR}
% \end{figure}

\subsection{Vertex-centered or cell-centered?}
\label{sec4.1}
In this work, we adopt a staggered-grid configuration, where the grid solutions are stored at cell centers rather than vertices.
We refer to these layouts as cell-centered and vertex-centered grids, respectively.
However, unlike finite-volume methods, the grid solutions at cell centers remain treated as pointwise values instead of cell averages.
While such a staggered arrangement has been previously used in the WCNS schemes to address corner issues in multi-domain computations~\cite{dengDevelopingHighOrderWeighted2000}, its explicit integration into finite-difference AMR frameworks has, to our knowledge, not been previously documented.

\begin{figure}[htpb]
    \centering

    % ---- 子图 A ----
    \begin{subfigure}[b]{0.48\linewidth}
        \centering
        \includegraphics[height=0.6\linewidth]{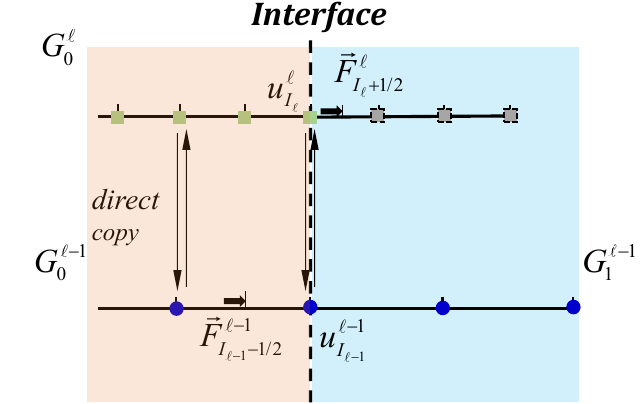}
        \caption{Vertex-centered scheme}
        \label{fig:data:a}
    \end{subfigure}
    \hfill
    % ---- 子图 B ----
    \begin{subfigure}[b]{0.48\linewidth}
        \centering
        \includegraphics[height=0.6\linewidth]{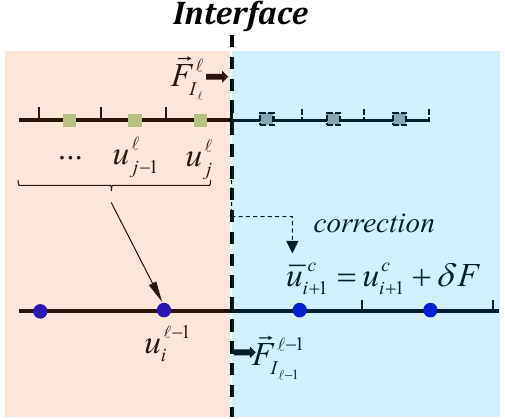}
        \caption{Cell-centered scheme}
        \label{fig:data:b}
    \end{subfigure}

    \caption{Different setups for high-order schemes in an AMR grid for a one-dimensional case. Fine and coarse solution points are illustrated by the green squares and blue circles, respectively, while the gray squares with dotted edges represent the ghost solution points of the fine grid.}
    \label{fig:data}
\end{figure}

To begin with, we consider the most common circumstance, i.e., a two-level AMR grid where a coarse-fine interface is contained therein.
The symbols $G^{\ell-1}_{m}, G^{\ell}_{n}$ denote the numerical solution on the coarse and fine domains, respectively, with the subscript $m$ indicating the index of a specific grid patch. 
The fundamental distinction between the two approaches lies that in the cell-centered approach the coarse and fine grid solutions $G^{\ell-1}_{m}, G^{\ell}_{n}$ are spatially non-overlapping, whereas the vertex-centered approach collocates the coarse grids with the finer ones.
This is illustrated in Fig.~\ref{fig:data}(a) and (b) in a one-dimensional computational space, while it extends naturally to higher dimensions.
Meanwhile, the coarse and fine grid patches are conformed at the flux points (denoted by $x^{\ell}_{j+1/2}$ or $x^{\ell-1}_{i-1/2}$) for the cell-centered scheme while at the solution points (denoted by $x^{\ell}_{j}$ or $x^{\ell-1}_{i}$) for the vertex-centered scheme. 

Here, we divide the solutions on a grid patch into interior and boundary solutions, respectively.
In vertex-centered grids, when the information flows in a direction from the fine to the coarse grids, the coarse solutions that are overlapped by the finer solutions in the interior region and at the boundaries are both updated by direct value replacement.
In return, when the coarse information is transferred to the fine patches, the additional non-collocated fine grid solutions are filled by interpolation on the coarse solutions.
It is easily observed that although this method is straightforward to implement and is reported to preserve the formal convergence order, it sacrifices the conservation inherently.
Moreover, by placing solution points (rather than flux points) at the end of each domain, the flux consistency at the boundaries is violated inevitably, which further exacerbates conservation loss between different AMR levels.

Specifically, at the coarse-fine boundary $x_I$ in Fig.~\ref{fig:data}a, as long as a pointwise replacement is employed between the coarse grid solution $u_{I_{\ell-1}}^{\ell-1}$ and the fine counterpart $u_{I_{\ell}}^{\ell}$, the respective numerical fluxes flowing into/out of the coarse/fine patches, denoted as $\vec{F}_{I_{\ell-1}-1/2}^{\ell-1}$ and $\vec{F}_{I_{\ell}+1/2}^{\ell}$ in this case, normally mismatches.
While such errors may be acceptable for smooth solutions, they can be particularly severe when a discontinuity is present at the coarse cells or crosses the coarse-fine interfaces, as reported in previous studies~\cite{sebastianMultidomainWENOFinite2003,shenAdaptiveMeshRefinement2011a,tangNonconservativeAlgorithmsGrid1999}.

Consequently, we choose the cell-centered method in Fig.~\ref{fig:data}(b), where it avoids direct overlapping of coarse and fine grid solutions within a same domain, and conforms different coarse and fine subdomains at the flux points instead.
This design enables the inter-patch conservation to be rigorously enforced by using a flux correction algorithm~\cite{wangEfficientGPUacceleratedAdaptive} where the coarse grid solution is corrected by an equivalent flux difference $\delta F_{I+1/2}$ at the end of a coarse step, which will be detailed in Section~\ref{sec4.3}.
For interior grid points, the non-collocated hierarchical structure in AMR grids also provides large flexibility that one can either use a high-order pointwise interpolation to maintain the formal convergence order or choose a conservative updating.
In this work, we apply the hybrid interpolation strategy for the first time in AMR, aiming to balance both accuracy and robustness.
The details of this strategy are presented in Section~\ref{sec5}.

\subsection{High-order space-time interpolation}
\label{sec4.2}

Based on the fifth-order WENO interpolation methodology presented in~\cite{sebastianMultidomainWENOFinite2003}, we re-derived the optimal linear weights, the coefficients for constructing candidate stencils, and the coefficients for smoothness indicator calculations for refinement ratios of 2 and 4, respectively, due to our adoption of a cell-centered rather than a vertex-centered solution point arrangement. 
Due to the non-collocated coarse and fine solution points, we also propose a high-order interpolation scheme for the restriction step based upon a six-point central stencil. 
For temporal accuracy, a high-order interpolation scheme adopted from~\cite{mccorquodaleHighorderFinitevolumeMethod2011} is applied.

Overall, this spatial and temporal interpolation strategy forms the foundation of our high-order staggered AMR-WENO framework, providing a robust and accurate solution pathway that markedly outperforms traditional vertex-centered approaches in compressible turbulence simulations.
The details for prolongation and restriction are given subsequently in this section.
In Section~\ref{sec5}, we further extend this AMR solver to more complex multiphysics problems in the presence of strong discontinuities.

\subsubsection{Prolongation}
\label{sec4.2.1}
\begin{figure}[htbp]
    \centering
    \includegraphics[width=0.9\linewidth]{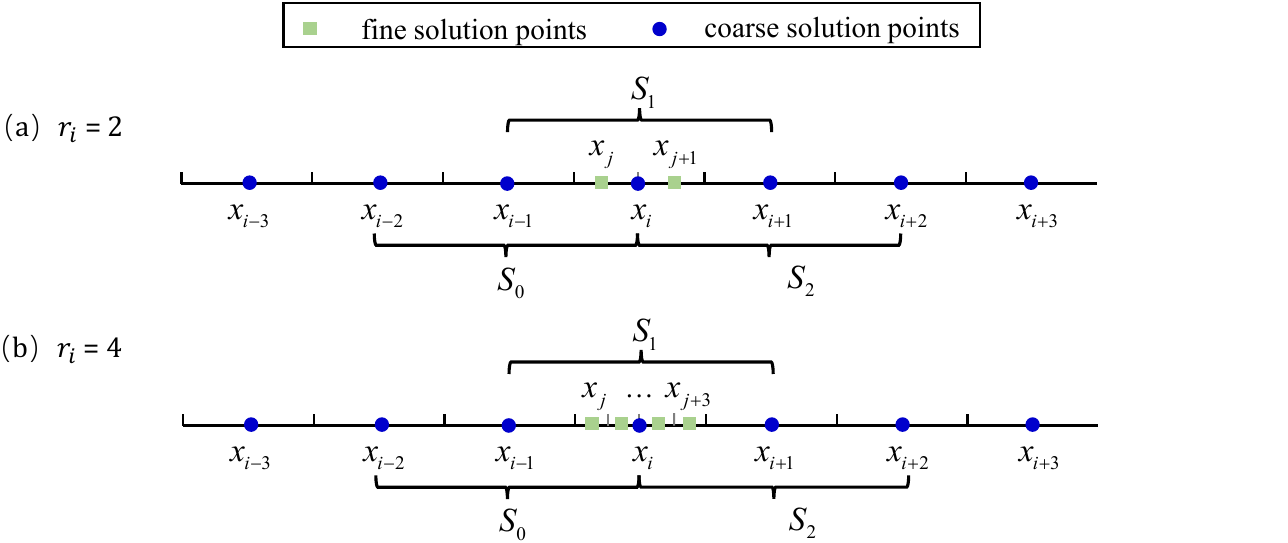}
    \caption{Fifth-order WENO interpolation schemes for staggered AMR grids during prolongation for a refinement ratio of (a) 2 and (b) 4. Fine and coarse solution points are shown as green squares and blue circles, respectively, while flux points are indicated by minor ticks along the axis.}
    \label{fig:prolong}
\end{figure}

Extended from~\cite{sebastianMultidomainWENOFinite2003}, we present the method for constructing a family of high-order WENO interpolation schemes for staggered AMR grids.
Figure~\ref{fig:prolong} shows the interpolation schemes during prolongation for refinement ratios $r_i$ of 2 and 4, respectively.
Different from~\cite{sebastianMultidomainWENOFinite2003}, each coarse solution is now located at the cell center $x_i$, which is subdivided into $r_i$ fine cells represented by $x_{j+s}, s = 0, ..., r_i-1$.
We begin with the case of $s = 0$ for $r_i=2$ and $s = 0,1$ for $r_i=4$, which corresponds to fine grid solutions positioned to the left of the coarse point, while the solution formulations on the right can be determined by symmetry.
To construct high-order interpolation for prolongation, an odd ($2r-1$)-large stencil centered at the coarse solution $x_{i}$ (i.e., $L_i = \{x_{i-r+1}, x_{i-r+2}, ..., x_{i+r-1}\}$) is employed, where $r$ is the width of candidate stencils.
In this case, $r=3$ sufficiently matches the main discretization schemes.

Similar to~\cite{sebastianMultidomainWENOFinite2003}, each $r$-point substencils, denoted as $S_k, k=0,\dots, r$, constructs a local Lagrange interpolant.
The associated coefficients $c_l^{(k)}$ for the Lagrange basis polynomials are listed in Table~\ref{tab:lagrange}, for a refinement ratio of 2.
Then the local interpolants are combined nonlinearly to a high-order approximation, with the corresponding optimal linear weights $d_k$ and smoothness indicators $\beta_k$ provided in Table~\ref{tab:weights_r2} and~\ref{tab5}, respectively, where $\beta_k$ is defined over an interval that contains the target interpolation point.
Here, we take the integration interval as $[x_{i-1/2}, x_{i+1/2}]$.

\begin{table}[!htbp]
\centering
\caption{Lagrange interpolation coefficients $c^{(k)}_l$ and optimal linear weights $d_k$ for $s = 0$ under a refinement ratio of 2.}
\footnotesize
\begin{subtable}[t]{0.48\textwidth}
    \centering
    \caption{Lagrange interpolation coefficients $c^{(k)}_l$.}
    \begin{tabular}{cccc}
        \toprule
        $k$ & $l = 0$ & $l = 1$ & $l = 2$ \\
        \midrule
        0 & 12/32  & -33/16  & 77/32 \\
        1 & -3/32  & 7/16    & 21/32 \\
        2 & 5/32   & 15/16   & -3/32 \\
        3 & 45/32  & -9/16   & 5/32 \\
        \bottomrule
    \end{tabular}
    \label{tab:lagrange}
\end{subtable}
\hfill
\begin{subtable}[t]{0.48\textwidth}
    \centering
    \caption{Optimal linear weights $d_k$.}
    \begin{tabular}{ccccc}
        \toprule
        $k$ & 0 & 1 & 2 & 3 \\
        \midrule
        $d_k$ & 3/256 & 99/256 & 693/1280 & 77/1280 \\
        \bottomrule
    \end{tabular}
    \label{tab:weights_r2}
\end{subtable}
\label{tab:combined}
\end{table}

\begin{table}[!htbp]
\centering
\caption{Smoothness indicator coefficients $b_{ij}^{(k)}$ for $s = 0$ under a refinement ratio of 2, with $i,j = 0,1,2$.}
\footnotesize
\begin{tabular}{ccccccc}
\toprule
$k$ & $u_{r-k+2} u_{r-k+1}$ & $u_{r-k+1} u_{r-k}$ & $u_{r-k} u_{r-k+2}$ & $u_{r-k}^2$ & $u_{r-k+1}^2$ & $u_{r-k+2}^2$ \\
\midrule
0 & $29/3$  & $-49/3$ & $-73/3$ & $10/3$ & $61/3$ & $22/3$ \\
1 & $11/3$  & $-19/3$ & $-31/3$ & $4/3$  & $25/3$ & $10/3$ \\
2 & $5/3$   & $-13/3$ & $-18/3$ & $4/3$  & $18/3$ & $4/3$  \\
3 & $11/3$  & $-31/3$ & $-19/3$ & $10/3$ & $25/3$ & $4/3$  \\
\bottomrule
\end{tabular}
\label{tab5}
\end{table}

For a mesh refinement ratio of 4, each coarse cell is subdivided into four fine cells located at $x_{j+s}$, with $s = 0,\dots,3$.
In this case, the stencil structure and interpolation coefficients for $s = 0,1$ are derived in a manner analogous to that for $r_i=2$.
Similarly, the cases $s = 2,3$ can be obtained by reflecting the results regarding the centered point at $x_i$.
The smoothness indicators used for $s = 0$ for $r_i=2$ still apply here, while the Lagrange coefficients and linear weights for the candidate stencils are summarized in Table~\ref{tab:combined_ref4}.

\begin{table}[!htbp]
\centering
\caption{Lagrange interpolation coefficients $c^{(k)}_l$ and optimal linear weights $d_k$ for $s = 0,1$ under a refinement ratio of 4.}
\footnotesize
% ---------- 子表 1：Lagrange coefficients ----------
\begin{subtable}[t]{0.55\textwidth}
    \centering
    \caption{Lagrange interpolation coefficients $c^{(k)}_l$.}
    \begin{tabular}{ccccccc}
        \toprule
         & \multicolumn{3}{c}{$s = 0$} & \multicolumn{3}{c}{$s = 1$} \\
        \cmidrule(r){2-4} \cmidrule(r){5-7}
        $k$ & $l = 0$ & $l = 1$ & $l = 2$ & $l = 0$ & $l = 1$ & $l = 2$ \\
        \midrule
        0 & -15/128 & 39/64  & 65/128  & -7/128  & 15/64  & 105/128 \\
        1 & 33/128  & 55/64  & -15/128 & 9/128   & 63/64  & -7/128  \\
        2 & 209/128 & -57/64 & 33/128  & 153/128 & -17/64 & 9/128   \\
        \bottomrule
    \end{tabular}
    \label{tab:lagrange_ref4}
\end{subtable}
\hfill
% ---------- 子表 2：WENO weights ----------
\begin{subtable}[t]{0.35\textwidth}
    \centering
    \caption{Optimal linear weights $d_k$.}
    \begin{tabular}{cccc}
        \toprule
        $s$ & $k = 0$ & $k = 1$ & $k = 2$ \\
        \midrule
        0 & 209/768 & 247/384 & 65/768 \\
        1 & 51/256  & 85/128  & 35/256 \\
        \bottomrule
    \end{tabular}
    \label{tab:weights_ref4}
\end{subtable}

\label{tab:combined_ref4}
\end{table}

This WENO interpolation approach yields formally fourth-order of accuracy for the prolongation step, which sufficiently satisfy the overall fifth-order spatial distretization schemes.
As observed, access to three additional coarse cells on either side of the base cell $x_i$ is required to form the full stencil.
While such extended stencils could potentially pose issues near the boundaries, fortunately, we numerically found that the proper nesting condition enforced in our AMR framework guarantees their availability.
As such, the required interpolation stencils are consistently well-posed throughout the entire fine-grid generation.

\subsubsection{Restriction}
\label{sec4.2.2}
In a cell-centered framework with even refinement ratios, the coarse grid solution points are non-overlapped with the finer ones, which requires additional interpolation from fine to coarse solutions to retain high-order.

\begin{figure}[H]
    \centering
    \includegraphics[width=0.9\linewidth]{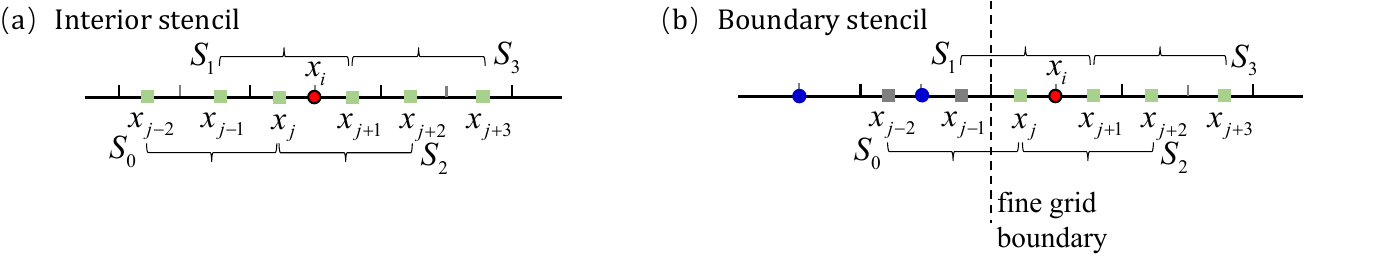}
    \caption{A blended WENO/ENO interpolation schemes for staggered AMR grids during restriction for (a) interior points and (b) points near the boundaries. The red circle denotes the target coarse grid point to be updated. Green and gray squares represent the valid and ghost fine grid solution points used for construction the candidate stencils, respectively. Blue circles indicate neighboring coarse grid points that provide additional information near the boundary.}
    \label{fig:restrict}
\end{figure}

In~\cite{zieglerAdaptiveHighorderHybrid2011}, Ziegler et al. adopted a simple Lagrange interpolation scheme for restriction.
However, our numerical experiments indicate that, although this approach is computationally efficient and straightforward to implement, it may lead to instabilities, especially in complex multiphysics scenarios.
To address this issue, we adopt an optimally sixth-order central WENO interpolation scheme for updating the coarse-grid solutions, which is summarized in~\ref{app1}.
In general, the methodology we employ here shares the same basic principles as that used for prolongation, but with a different number of stencil points for constructing the high-order polynomial approximations.
It is worth emphasizing that, unlike conventional WENO reconstructions that operate on flux vectors and must account for upwind characteristics, our procedure directly interpolates the conservative solution variables.
As a result, directional bias is not required in this step.
% Although our background scheme shares similarities with the previous WENO-CU6 formulation~\cite{huAdaptiveCentralupwindWeighted2010}, the key difference lies that in WENO-CU6, the weighting of the downstream stencil is evaluated over the entire large stencil,
% whereas in our WENO interpolation, the smoothness indicators are evaluated only within the local three-point candidate stencils.

The interpolation schemes for the interior and boundary coarse-grid solutions are illustrated in Fig.~\ref{fig:restrict}, respectively.
For interior coarse-grid points, accessing the three nearest fine-grid points on each side is straightforward.
However, for points located near the fine-grid boundaries, the original large stencil is truncated because of insufficient stencil points.
To avoid order reduction, the existing coarse-grid solutions outside the fine region are incorporated to construct a complete stencil.
In practice, two fine-grid ghost cells are introduced to form a six-point central WENO stencil, as illustrated in Fig.\ref{fig:restrict}b.
The ghost-cell values are filled using a local interpolation technique similar to prolongation, based on the surrounding coarse-grid solutions.
Our numerical experiments demonstrate that this approach provides accurate and stable solutions, except when a discontinuity is located near the boundary.
However, in such cases, we can resort to a low-order conservative interpolation scheme, which will be described in detail in Section\ref{sec5}.

In summary, our restriction strategy is capable of maintaining high-order accuracy in smooth regions.
Although this interpolation scheme may encounter robustness issues near the boundaries due to repeated interpolations, in extreme conditions we can switch to a conservative averaging at shock discontinuities.
Its corresponding coefficients $c^{(k)}_l, d_k$ and $b_{ij}^{(k)}$ are given in Tables~\ref{tab:combined2}.

% \begin{figure}[htpb]
%     \centering
%     \includegraphics[width=0.4\linewidth]{figs/restriction2.pdf}
%     \caption{Illustration of constructing a central stencil $L_i$ by supplying ghost cells at the fine level versus local truncation of the stencil $L_i'$ near the boundaries.}
%     \label{fig:truncated}
% \end{figure}

\begin{table}[htbp]
\centering
\caption{Coefficients, optimal linear weights, and smoothness indicator coefficients used in central WENO interpolation during restriction.}
\footnotesize

% ---------- 第一行：子表 1 和 子表 2 ----------
\begin{subtable}[t]{0.45\textwidth}
    \centering
    \caption{Local interpolation coefficients $c^{(k)}_l$.}
    \begin{tabular}{cccc}
        \toprule
        $k$ & $l = 0$ & $l = 1$ & $l = 2$ \\
        \midrule
        0 & 3/8  & -5/4  & 15/8 \\
        1 & -1/8 & 3/4   & 3/8 \\
        2 & 3/8  & 3/4   & -1/8 \\
        3 & 15/8 & -5/4  & 3/8 \\
        \bottomrule
    \end{tabular}
    \label{tab:coeff}
\end{subtable}
\hfill
\begin{subtable}[t]{0.45\textwidth}
    \centering
    \caption{Optimal linear weights $d_k$.}
    \begin{tabular}{ccccc}
        \toprule
         & $k = 0$ & $k = 1$ & $k = 2$ & $k = 3$ \\
        \midrule
        $d_k$ & 1/32 & 15/32 & 15/32 & 1/32 \\
        \bottomrule
    \end{tabular}
    \label{tab:weights}
\end{subtable}

\vspace{1em} % 第一行与第二行之间的垂直间距，可根据需要调整

% ---------- 第二行：子表 3 ----------
\begin{subtable}[t]{\textwidth}
    \centering
    \caption{Smoothness indicator coefficients $b_{ij}^{(k)}$.}
    \begin{tabular}{ccccccc}
        \toprule
        $k$ & $u_{i-r+k}u_{i-r+k+2}$ & $u_{i-r+k}u_{i-r+k+1}$ & $u_{i-r+k+1}u_{i-r+k+2}$ & $u_{i-r+k}^2$ & $u_{i-r+k+1}^2$ & $u_{i-r+k+2}^2$ \\
        \midrule
        0 & 37/6 & -31/3 & -49/3 & 25/12 & 40/3 & 61/12 \\
        1 & 13/6 & -13/3 & -19/3 & 13/12 & 16/3 & 25/12 \\
        2 & 13/6 & -19/3 & -13/3 & 25/12 & 16/3 & 13/12 \\
        3 & 37/6 & -49/3 & -31/3 & 61/12 & 40/3 & 25/12 \\
        \bottomrule
    \end{tabular}
    \label{tab:smoothness}
\end{subtable}

\label{tab:combined2}
\end{table}

\subsubsection{Boundary conditions on fine grids}
\label{sec4.2.3}
To maintain high-order accuracy, ghost-cell interpolations on the fine grids must also be performed to provide accurate boundary conditions before advancing the solution in time.
In the subcycling time-stepping paradigm, because of temporal refinement, the ghost-cell values for the fine-grid solutions at intermediate time levels need to be interpolated.
In the original finite-volume AMR method, a second-order linear interpolation scheme is typically employed for this purpose.
However, in a high-order AMR-WENO framework, the temporal interpolation must be consistent with the temporal integration method.
Otherwise, the overall temporal accuracy will deteriorate, leading to a loss of fidelity in unsteady flow simulations.
The entire procedure consists of two main components:
\paragraph{Spatial interpolation}
\label{sec4.2.3.1}
The spatial interpolation procedure essentially follows the same strategy as that used in the prolongation step.
It is important to note that this process does not introduce any loss of conservation, even though a non-conservative interpolation is employed.
This is because the interpolation is only applied to populate ghost cells, rather than to update the valid solution cells.
Consequently, a high-order WENO interpolation scheme is adopted in a dimension-by-dimension manner to ensure accuracy and robustness.
% In contrast, Lagrange interpolation has been numerically tested and found to exhibit stability issues when applied to complex problems.

\paragraph{Temporal interpolation}
\label{sec4.2.3.2}
Due to the use of subcycling time-stepping approach, the intermediate states on the fine grids must be interpolated from its overlaid coarse grid solutions.
To ensure the accuracy of the time interpolation between different AMR levels, we adapt the method presented by McCorquodale and Colella~\cite{mccorquodaleHighorderFinitevolumeMethod2011} to be applied to the third-order Runge--Kutta time method.
For a time ratio $\eta$, with $\eta \in [0, 1]$, an intermediate state is interpolated using
\begin{equation}
\label{RK3Interp}
u\left(t^{n}+\eta \Delta t_{c}\right)=u^{n}+k_{1} \eta+\left(-\frac{5}{6} k_{1}+\frac{1}{6} k_{2}+\frac{2}{3} k_{3}\right) \eta^{2}+O\left(\Delta t_{c}^{3}\right),
\end{equation}
where $u^n$ is the coarse solution at time $t^n$, and $k_1$, $k_2$, $k_3$ are spatial operators at different stages. 
The intermediate time solution at the refined time level $t_f$ that does not exist at its overlaid coarse level is obtained immediately by substituting $\eta=(t_f-t^n)/\Delta t_c$ into Eq.~\ref{RK3Interp}.

On the other hand, by substituting 
\begin{equation}
\label{eta}
    \eta_1=\frac{t^{f}+\Delta t_f-t^{n}}{\Delta t_c}, \quad \eta_2=\frac{t^{f}+\Delta t_f/2-t^{n}}{\Delta t_c}, 
\end{equation}
where $t^{f}$ is the current initial time at the fine level, we can also directly find the third-order approximations to the intermediate states $u^{(1)},u^{(2)}$ at the fine level.
However, as reported in~\cite{mccorquodaleHighorderFinitevolumeMethod2011}, a mismatch may occur between the interpolated values on the ghost cells and those on the interior cells. 
This discrepancy could potentially degrade the accuracy even for smooth structures, with the WENO interpolation procedure incorrectly detecting non-smoothness. 
To mitigate this issue, we perform a Taylor expansion in the interpolated solutions at $t^n$ and ensure them aligned with the stages of the third-order Runge–Kutta scheme. 
Further details of this approach can be found in~\cite{mccorquodaleHighorderFinitevolumeMethod2011}. 
By doing such, the interpolated values achieve third-order temporal accuracy and are fully compatible with the RK3 scheme. 
This temporal alignment avoids spurious accuracy degradation both spatially and temporally across different AMR levels.

\subsection{Flux correction}
\label{sec4.3}
In the presence of subcycling-in-time, each AMR level advances with its own time step individually and performs flux correction (refluxing) at the end of a coarse step to ensure conservation across coarse–fine interfaces. 
Without this correction, inconsistencies between fluxes on different levels may lead to violations of global conservation.

For explicit Runge-Kutta solvers, the flux differences at coarse–fine interfaces at each stages must be scaled and accumulated. For Runge–Kutta (RK) time-stepping methods in arbitrary orders, the semi-discrete formulation can be written as
\begin{equation}
U^{n+1} = U^{n} + h \sum_{m=1}^s W_m K_m ,
\end{equation}
where
\begin{equation}
\begin{aligned}
K_1 &= f\left(t_n, u(t_n)\right), \\
K_m &= f\left(t_n + h \alpha_{m}, \, u(t_n) + h \sum_{l=1}^{m-1} \beta_{ml} K_l \right), \quad m = 2, \dots, s ,
\end{aligned}
\end{equation}
In this formulation, the final conservative update incorporates contributions from both the local RK fluxes and the accumulated reflux terms. 
As an example, the flux correction at a coarse–fine interface with the normal direction aligned with the $y$-axis can be expressed as
\begin{equation}
    U_{i, j-1, k}^{l-1(m)}=U_{i, j-1, k}^{l-1(m)} + \sum_{m=1}^sW_{m} \delta \hat{f}_{i, j-1 / 2, k}^{l-1} \frac{\Delta t^{l-1}}{\Delta y^{l-1}}
\end{equation}
where $l$ is the finer level, $\hat{f}_{i, j-1 / 2}^{l-1}$ is the flux difference at the coarse-fine interfaces, whose expression can be found in~\cite{bergerLocalAdaptiveMesh1989, deiterdingAdaptiveMultiresolutionAdaptive2009}.
This procedure ensures that the fluxes computed on the fine and coarse levels are fully consistent at their shared interfaces, thereby preserving local and global conservation properties in subcycled AMR–WENO simulations.

\section{A Hybrid Interpolation Strategy in SAMR}
\label{sec5}
\subsection{Issues from non-conservative interpolation}
\label{sec5.1}
During AMR, cells tagged by the refinement criteria are typically surrounded by additional buffer zones, as shown in Fig.~\ref{fig:buffer}.
The purpose of the buffer zones is to ensure that important moving flow features, such as shocks or other sharp gradients, remain mostly contained within the refined region until the next grid adaptation step.
This setup is essential for high-order non-conservative interpolation schemes, such as WENO interpolation, which rely on the assumption that discontinuities are always resolved at the finest level. 
Under this ideal scenario, the WENO interpolation method was reported to yield satisfactory accuracy and stability, although the mass conservation is lost inevitably~\cite{shenAdaptiveMeshRefinement2011}.

However, the hypophysis is often too idealized for practical simulations.
In real applications, grid adaptation must be balanced with computational costs, especially in large-scale multiphysics simulations where frequent global regridding is prohibitively expensive.
As a result, there is always the possibility that a discontinuity may move partially or entirely out of the refined region before the next grid adaptation is performed.
When this occurs, a strictly non-conservative high-order interpolation becomes problematic: although it provides high accuracy in smooth regions, it fails to enforce conservation across coarse–fine interfaces and at the coarse-grid solutions, potentially leading to unphysical results in shocked flows.
On the other hand, a fully conservative interpolation approach ensures strict conservation but is limited to only second-order accuracy, which substantially smears out the high-wavenumber structures in smooth regions.

To better illustrate this issue, we identify three common situations for discontinuous flows when using high-order non-conservative interpolation:

\begin{figure}[htpb]
    \centering
    \includegraphics[width=0.5\linewidth]{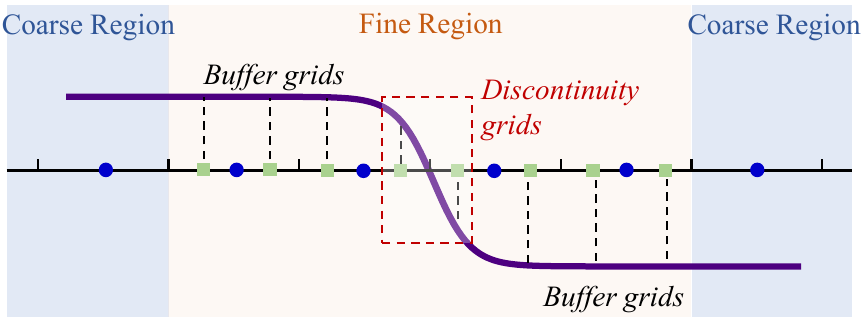}
    \caption{A typical grid-refinement case where a discontinuity is present with buffer cells surrounded.}
    \label{fig:buffer}
\end{figure}

\begin{figure}[htpb]
    \centering
    \includegraphics[width=0.99\linewidth]{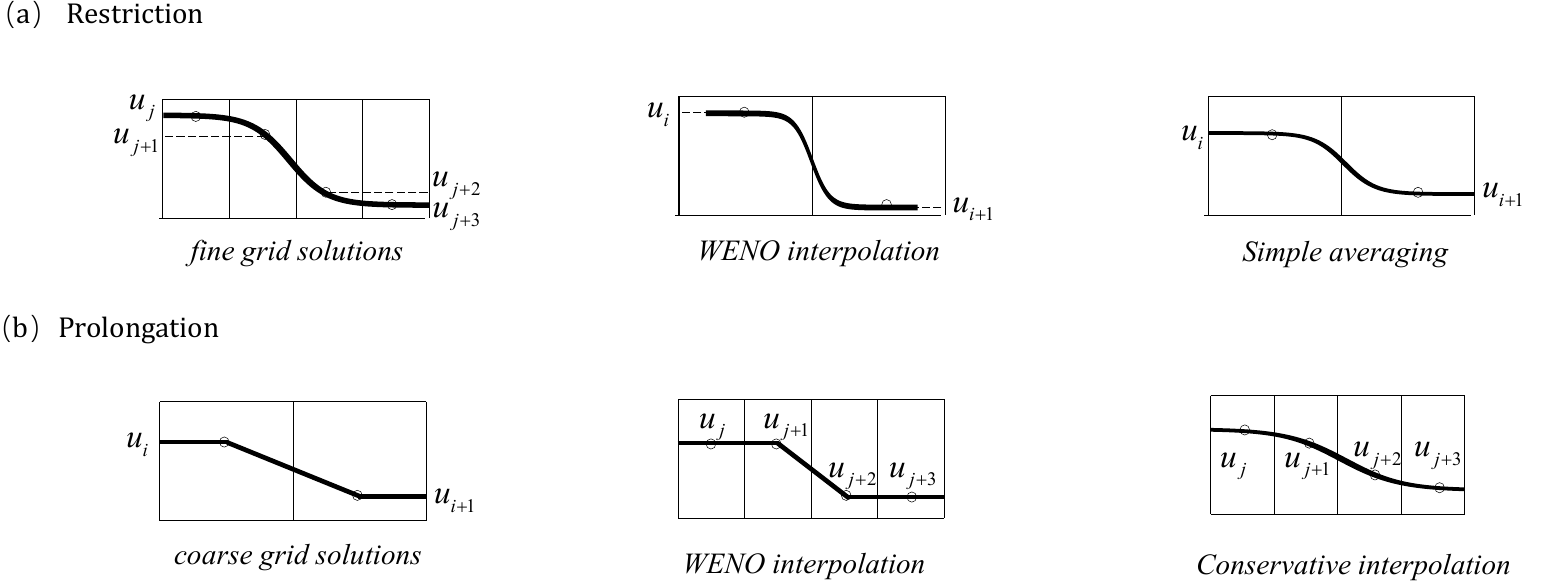}
    \caption{Illustration of the underlying mechanisms of shock-profile distortion by non-conservative interpolation method in the (a) restriction and (b) prolongation steps, as compared with the conservative counterpart.}
    \label{fig:flawed}
\end{figure}

\begin{enumerate}[label=\Roman*).]
    \item \label{item:1} \textbf{The discontinuity remains entirely within the finest grid region throughout the simulation.}

    In this idealized scenario, the discontinuity is always fully captured by the finest grid during the dynamic mesh refinement process.  
    Although the non-conservative interpolation used in the restriction step may introduce slight inconsistencies when projecting solutions from fine grids to coarse grids, these errors remain confined to the coarse level.  
    As long as the fine-grid solution is accurate, the Rankine--Hugoniot (R-H) relations at the discontinuity are essentially preserved, and the overall solution remains physically meaningful.

    Many previous AMR studies based on finite-difference WENO frameworks~\cite{shenAdaptiveMeshRefinement2011a,chenFifthorderFiniteDifference2016,wangParallelAdaptiveMesh2015a} have implicitly assumed this condition, thereby neglecting the conservation issue.  
    However, for practical AMR applications where robustness is critical, this assumption is often overly optimistic and difficult to maintain in real-world computations.
    
    \item \label{item:2} \textbf{The discontinuity crosses the coarse--fine interface.}

    When a discontinuity moves out of the refined region and across the boundary between coarse and fine grids, purely non-conservative method leads to artifacts.  
    As the discontinuity repeatedly crosses between coarse and fine regions during the dynamic refinement process, the conservation errors are not only introduced but also accumulated over time.  
    This cumulative effect is particularly detrimental, leading to significant deterioration in both the fidelity and robustness of the simulation.
    Fundamentally, this phenomenon can be interpreted as a violation of the R-H condition, in which the true coarse-grid state $u_i$ is perturbed by a conservation error $\delta u$ of order $O(\Delta x)$, leading to a spurious state $\overline{u}_i$ given by
    \begin{equation}
    \overline{u}_i = u_i + \delta u.
    \end{equation}
    Nevertheless, this scenario can be addressed perfected in this work following the prediction-correction method by employing a cell-centered schemes, where the pointwise values are corrected at the end of each coarse step with the flux differences.

    \item \label{item:3} \textbf{The discontinuity gets instantaneously refined or coarsen.}

    This is the most challenging situation in practical AMR simulations.  
    It occurs when the refinement criteria abruptly detects a discontinuity or the discontinuity is weakened as the time evolves.

    We have analyzed in~\ref{item:1} that the non-conservative interpolation method stands in scenario~\ref{item:1} because the conservation errors due to restriction step are refreshed at each step and are confined only at the coarser level.
    However, it becomes problematic when the erroneous values at the coarse levels are actually evolved with time instead.
    As illustrated in Fig.~\ref{fig:flawed}(a), consider a one-dimensional AMR grid where the discontinuity is initially resolved by four fine-grid points, labeled $u_j, u_{j+1}, u_{j+2}, u_{j+3}$.  
    During the restriction step, the coarse-grid values are interpolated rather than averaged from the fine-grid values.  
    In a WENO interpolation scheme, the algorithm naturally selects the smoothest local stencil, causing the interpolated coarse values $u_i$ and $u_{i+1}$ to approach the extreme values on either side of the discontinuity.  
    This process distorts the original R-H relations, leading to an artificial steepening of the discontinuity(see Sections~\ref{sec6.3.1} and~\ref{sec6.3.2}).
    In contrast, the simple averaging approach yields an excellent conservative outcome, thereby accurately preserving the original R-H relations.

    This scenario differs fundamentally from scenario~\ref{item:1}: here, the corrupted coarse values are subsequently used as the reconstruction stencil ones for time integration.  
    As a result, the local conservation errors are no longer confined locally
    Instead, they are propagated to adjacent coarse grids.  
    This issue is clearly demonstrated 
    
    The issue is exacerbated when combined with scenario~\ref{item:2}: in our numerical experiments (see Section~\ref{sec6.2}), non-conservative WENO interpolation leads to numerical breakdown when strong shocks traverse AMR coarse-fine interfaces.

    In contrary, when a discontinuity is abruptly recognized and refined, the non-conservative interpolation method yields erroneous steeping of the existing shock, which is illustrated in Fig.~\ref{fig:flawed}(b).

\end{enumerate}

In summary, it is found that no single interpolation scheme can simultaneously maintain high-order accuracy in smooth regions and guarantee conservation at discontinuities. 
The non-conservative high-order interpolation compromise the mass conservation, thereby restricted to apply only to ideal scenarios; low-order conservative method tends to loss accuracy on smooth regions, and thus deteriorate its main discretization schemes on resolving high-wavenumber structures.
Although it has been proven the conservative error by WENO interpolation converges at the order of $O(\Delta x)$ in shocked regions~\cite{sebastianMultidomainWENOFinite2003,tangNonconservativeAlgorithmsGrid1999}, the flaws could continuously accumulate at the AMR interfaces or during the long-time computation, which is disadvantageous for high-fidelity multiscale shock-turbulence simulations.

Therefore, we present here to use a hybrid interpolation strategy that combines a \textit{high-accuracy} non-conservative interpolation on smooth regions and a \textit{low-order} conservative interpolation scheme to enforcing conservation at discontinuities.
The interpolation schemes are switched through a reliable shock detection method that can accurately flag the shocks, as previously used for developing low-dissipating hybrid numerical schemes for turbulence modeling (e.g., ~\cite{adamsHighResolutionHybridCompactENO1996,pirozzoliConservativeHybridCompactWENO2002,zieglerAdaptiveHighorderHybrid2011}).

\subsection{Troubled-cell detection}
\label{sec5.2}
It is claimed that any means to preserve the conservation will reduce the schemes to no more than second order of accuracy~\cite{sebastianMultidomainWENOFinite2003}.
In this work, we found it applies to either flux correction or conservative interpolation processes.
Therefore, in this paper, we present to use troubled-cell detection method for distinguish shocks and smooth regions in the flow field.
In smooth regions, we use a pointwise high-order WENO interpolation scheme to ensure the overall accuracy; while at the troubled cells, we switch to a conservative interpolation along with a flux correction step to preserve the local conservation.
This method effectively eliminates the conservator errors that could occur at the shock discontinuities, avoiding potential numerical artifacts on producing an nonphysical shock speed.

In previous literature~\cite{wang2025adaptive,pantanoLowNumericalDissipation2007}, a simple detection criteria given by normalized curvature of density or pressure was widely used.
On one hand, it is straightforward to implement and well-compatible with the local grid refinement of AMR.
On the other hand, this method underperforms on cells with relatively coarse spatial steps where a smooth flow might be recognized as an initial discontinuity thereby its flexibility is quite limited.
Some attempts by construction WENO smoothness indicators have also been tried; but again, the same issue applies and thus is not widely employed for its increased complexity.
In this work, we employed an alternative method, by leveraging an approximate Riemann solver while combining a mapped pressure curvature to enhance efficiency and stability, as similarly done in ~\cite{zieglerAdaptiveHighorderHybrid2011}.

To state here,  a shock is detected if the intermediate states satisfy the condition
\begin{equation}
    \left|u_{R} \pm a_{R}\right|<\left|u_{*} \pm a_{*}\right|<\left|u_{L} \pm a_{L}\right|, \quad 
    \theta_{i} > \theta_{map}.
\end{equation}
where the intermediate states are computed as the Roe averages by
\begin{equation}
    u_{*}=\frac{\sqrt{\rho_{L} u_{L}}+\sqrt{\rho_{R} u_{R}}}{\sqrt{\rho_{L}}+\sqrt{\rho_{R}}}, \quad 
    h_{*}=\frac{\sqrt{\rho_{L} h_{L}}+\sqrt{\rho_{R} h_{R}}}{\sqrt{\rho_{L}}+\sqrt{\rho_{R}}}, \quad
    a_{*}=\sqrt{\left(\gamma-1\right)\left(h_{*}-\frac{1}{2} u_{*}^{2}\right)},
\end{equation}
and the mapped pressure curvature defined as
\begin{equation}
    \theta_{i}=\frac{2 S_{i}}{\left(1+S_{i}\right)^{2}} \quad \text { with } S_{i}=\frac{\left|p_{i-1}-2 p_{i}+p_{i+1}\right|}{p_{i-1}+2 p_{i}+p_{i+1}} .
\end{equation}

A threshold value of normalized Lax residual, $\alpha_{\text{Liu}}/\alpha$, is used additionally for elimination of weak compression waves, which are adequately handled by the high-order interpolation scheme without oscillations.
According to our experiments, this troubled-cell detection method is much more stable and insensitive to small perturbations.

Additionally, we also found that the high-order WENO interpolation is significantly more robust than the Lagrange interpolation of comparable order, and is therefore adopted in this work. 
For conservative interpolation, a linear spatial and temporal scheme is employed to achieve second-order accuracy by
\begin{equation}
    u(x, y, z) = u_{i_c,j_c,k_c} + s_x \cdot \Delta x + s_y \cdot \Delta y + s_z \cdot \Delta z,
\end{equation}
\begin{equation}
    u(x,y,z,t) = u^n(x,y,z) + \frac{t - t^n}{\Delta t_c}[u^{n+1}(x,y,z)-u^n(x,y,z)]
\end{equation}
where $s_x, s_y, s_z$ denote the spatial slopes, $\Delta x, \Delta y, \Delta z$ are the spatial offsets, and $t$ is the target time. 
The spatial slopes are further constrained using a local Lax–Friedrichs limiter to ensure numerical stability.

This hybrid interpolation framework provides both high flexibility and accuracy, as demonstrated in canonical test problems. 
The only drawback is that it requires an explicit threshold to determine when to switch interpolation schemes as well as using the flux correction, which is a common issue encountered in other hybrid approaches. 
In this work, we recommend using $\alpha_{\text{Liu}}/\alpha = 0.1$ and $\theta_{\text{map}} = 10^{-2}$.

% \subsection{Sensitivity analysis}
% brief summary (will be detailed later):
% ...

% \section{GPU Strategy}
% \label{sec}
% brief summary (will be detailed later):
% portable with the use of C++ macros and GPU extended lambdas, mpi+CUDA, allocate initial device memory pool with minimized memory movement,  parallelize over a valid box at a time, use of cuda stream for high concurrency, 

% In AMR, mesh generation, logical computation handled on host while float-point computation resides on device

\section{Results}
\label{sec6}
\subsection{Convergence results}
\label{sec6.1}
First, we consider the 1-D steady Navier-Stokes equations to confirm the spatial convergence of the AMR solution.
To begin with, evaluation of the $L_1$ norm on an AMR solution can be expressed as
\begin{equation}
    L_{1}(q)=L_{1}^{l_{f}}\left(\Delta x_{k}^{l_{f}},G^{l_{f}}\right)+\sum_{k=0}^{l_{f}-1} L_{1}\left(\Delta x_{k}^k, G^{k} \backslash G^{k+1}\right)
\end{equation}
where $l_{f}$ is the finest level, $G^{k} \backslash G^{k+1}$ is the grid region on level $k$ that is non-overlapped by level $k+1$. 
This avoids repetitive counting for fine-grid error contributions.

The computational domain of $[0,1]$ is divided into $N$ cells; in this case, we refine $x\in[0,0.5]$ by a refinement ratio of 2 in a 2-level AMR configuration.
Refinement ratios above 2 together with higher-level refinement were also tested to yield similar results and thus will not be shown here.
The method of manufactured solutions (MMS)~\cite{malayaMASALibraryVerification2013} is employed, in which an analytical source term is added to the governing equations to construct a prescribed Navier–Stokes solution.
It is shown in Table~\ref{tab:steady_NS} that the results converge to the formal accuracy of the spatial discretization scheme, demonstrating the validness of the spatial WENO interpolation during AMR prolongation and restriction steps.

\begin{table}[!htbp]
\centering
\caption{Convergence results for one-dimensional steady Navier–Stokes equations and unsteady Euler equations, respectively.}
\footnotesize
% ---------- 子表 1：Steady Navier-Stokes ----------
\begin{subtable}[t]{0.63\textwidth}
    \centering
    \caption{Steady Navier–Stokes equations at $t$ = 0.05}
    \begin{tabular}{ccccccc}
        \toprule
        $N$ & $L_1$ error of $\rho$ & Order & $L_1$ error of $u$ & Order & $L_1$ error of $p$ & Order \\
        \midrule
        32  & 2.18E-7  & --    & 7.34E-6  & --    & 6.81E-3  & --    \\
        64  & 1.06E-8  & 4.35  & 2.67E-7  & 4.78  & 3.15E-4  & 4.43  \\
        128 & 3.65E-10 & 4.86  & 8.18E-9  & 5.02  & 1.05E-5  & 4.90  \\
        256 & 8.62E-12 & 5.40  & 2.53E-10 & 5.01  & 3.45E-7  & 4.93  \\
        \bottomrule
    \end{tabular}
    \label{tab:steady_NS}
\end{subtable}
\hfill
% ---------- 子表 2：Unsteady Euler ----------
\begin{subtable}[t]{0.30\textwidth}
    \centering
    \caption{Unsteady Euler equations at $t = 1.0$}
    \begin{tabular}{ccc}
        \toprule
        $N$ & $L_1$ error of $\rho$ & Order \\
        \midrule
        32  & 1.61E-4  & --   \\
        64  & 6.40E-6  & 4.65 \\
        128 & 2.25E-7  & 4.83 \\
        256 & 7.44E-9  & 4.92 \\
        \bottomrule
    \end{tabular}
    \label{tab:unsteady_Euler}
\end{subtable}

\label{tab:combined_convergence}
\end{table}

Next, to verify the temporal accuracy on an unsteady problem, the one-dimensional unsteady Euler equations are solved.
Rather than testing the temporal convergence in isolation, we assess the overall error convergence of the coupled space–time scheme using a time step of $\delta t=0.2\Delta x^2$.
The initial condition is specified as a smooth sinusoidal profile
\begin{equation}
(\rho,u,p) = (1+0.2sin(2\pi x),1,1).
\end{equation}
A two-level grid hierarchy is employed, with the mesh dynamically refined according to the absolute value criterion $\rho>1.19$.
As shown in Table~\ref{tab:unsteady_Euler}, the density solution converges at the expected formal order of accuracy, clearly demonstrating the effectiveness of the high-order temporal interpolation method.

For this smooth problem, as shown in Table~\ref{tab:1d-ce-convergence}, the conservation error (CE) converges at fifth order, which is in consistency with the truncation error of the underlying WENO interpolation.
Unlike the multidomain WENO computation by Sebastian and Shu~\cite{sebastianMultidomainWENOFinite2003}, where the subdomain interfaces inherently introduce non-conservative errors of second-order convergence, our AMR framework maintains a fully nested hierarchy. 
In this smooth problem without flux correction, the prolongation and restriction steps by high-order interpolations yield conservation errors that converge to the formal order of the underlying WENO scheme (fifth order in our case). 
However, we also find that once flux correction is activated, the coarse-fine flux mismatch dominates and limits the global conservation error convergence to second order, consistent with the previous findings~\cite{sebastianMultidomainWENOFinite2003,tangNonconservativeAlgorithmsGrid1999}.

\begin{table}[htbp]
\centering
\caption{Convergence results of conservation error (CE).}
\footnotesize
\begin{tabular}{ccccccc}
\toprule
$N$ & $L_1$ error of $\rho$ & Order & $L_1$ error of $\rho u$ & Order & $L_1$ error of $\rho E$ & Order \\
\midrule
32   & 5.95E-4  & --   & 6.95E-4  & --   & 3.47E-4    & --     \\
64   & 2.13E-5  & 5.02 & 2.13E-5  & 5.02 & 1.06E-5    & 5.02 \\
128  & 5.03E-7  & 5.40 & 5.01E-7  & 5.40 & 2.50E-7    & 5.41 \\
256  & 1.20E-8  & 5.38 & 1.20E-8  & 5.38 & 1.00E-8    & 4.64 \\
\bottomrule
\end{tabular}
\label{tab:1d-ce-convergence}
\end{table}

\subsection{Shock–interface interaction problems}
\label{Sec6.2}
The problem with shock discontinuity passing from one subdomain to another~\cite{sebastianMultidomainWENOFinite2003} in multi-domain computation, or from one level to another in AMR calculation, has always been an intricate task that is under resolved in previous studies~\cite{sebastianMultidomainWENOFinite2003,shenAdaptiveMeshRefinement2011}.
In practical applications, this circumstance is usually encountered when a shock propagates from a finer patch to a coarser patch. 
In this section, three model problems of shock passing through the coarse-fine interface are tested with increasing challenges: a) isentropic Euler–Burgers coupling problem, b) slowly moving shock problem with moderate shock strength and c) slowly moving shock problem with strong shock strength.

The computational domain remains as $[0,1]$. 
For problem a), periodic conditions are imposed for both ends of the domain to allow multiple shock passes through the coarse-fine interfaces.
The exact condition adopts a velocity field prescribed by the inviscid Burgers solution with the density and pressure distributions follow the isentropic relations.
The initial condition is given by
\begin{equation}
    u(x, 0) = \frac{2}{\gamma+1} V(x), \quad 
    \rho(x) = {\rho_{\infty}}\left[1 + \frac{\gamma-1}{2}\left(\frac{u(x)}{a_{\infty}}\right)\right]^{\frac{2}{\gamma-1}}, \quad
    p(x) = {p_{\infty}}\left[1 + \frac{\gamma-1}{2}\left(\frac{u(x)}{a_{\infty}}\right)\right]^{\frac{2\gamma}{\gamma-1}}, \quad 
    V(x) = \frac{1}{2 \pi t_{s}} \sin (2 \pi x).
\end{equation}
The calculation lasts for $t=4t_s$, where $t_s=1.0$ is the time for shock formation. 
The other freestream parameters are set as $\rho_{\infty} = 0.1$, $p_{\infty} = 1.0$ and $a_{\infty}=\sqrt{\gamma p_{\infty}/\rho_{\infty}}$.

For problems b) and c), the exact pre- and post-shock conditions are applied at the right and left boundaries, respectively. 
In b), a Mach-2 shock, initially located at $x=0.05$, moves slowly towards the positive direction with a shock speed $s$ = 0.01, where the initial shock condition is
\begin{equation}
    (\rho, u, p)=\left\{\begin{array}{ll}(2.6667,-0.8774,4.5), & x \in[0,0.05) \\(1.0,-2.3564,1.0), & x \in[0.05,1.0] \end{array}\right.
\end{equation}
The calculation ends at $t = 90$ when the shock arrives ar $x=0.95$.

In problem c), the exact solution is a Mach-3 shock initially situated at $x=0.37$ and moving with a same shock speed $s$ = 0.01.
The initial condition is
\begin{equation}
    (\rho, u, p)=\left\{\begin{array}{ll}(3.86, -0.91,10.33), & x \in[0,0.37) \\(1.0,-3.54,1.0), & x \in[0.37,1.0] \end{array}\right.
\end{equation}
The calculation ends at $t = 30$ where the exact final location of the shock is $x=0.67$.

\begin{figure}[htpb]
    \centering
    \includegraphics[width=0.9\linewidth]{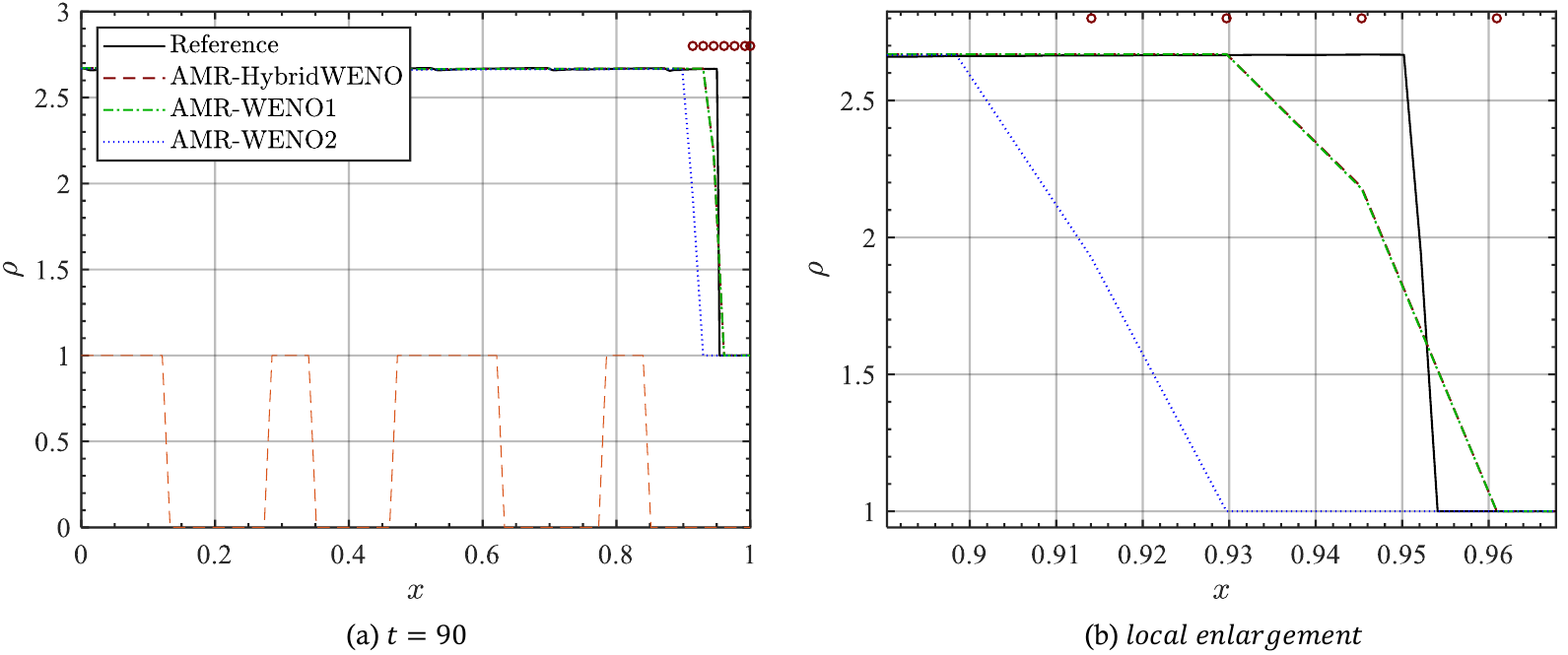}
    \caption{The slowly moving shock problem at a 2-level fixed AMR grid with a shock strength $Ma_s$ = 2. The AMR-HybridWENO, AMR-WENO1 and AMR-WENO2 solutions are obtained by hybrid interpolation, WENO interpolation with and without flux correction, respectively. The red circles indicates the troubled cell locations in AMR-HybridWENO method. The orange dashed line indicates the refinement level at the current AMR hierarchy.}
    \label{fig:ma2-shock}
\end{figure}

For all three benchmark problems, a uniform grid with \( N = 512 \) cells is employed to obtain the reference solution. 
For the AMR configuration of all benchmarks, a thin sheet of each subdomain at a width of 1/10 is tagged for refinement, to ensure the presence of sufficient coarse–fine interfaces. 
After the grid-merging procedure, a total of seven coarse–fine interfaces are formed eventually, as illustrated in Fig.~\ref{fig:ma2-shock}. 
In AMR, the finest grid resolution is set to be half of the reference resolution, with a base grid of \( N = 64 \) and a refinement ratio of 2.

\begin{figure}[htpb]
    \centering
    \includegraphics[width=0.9\linewidth]{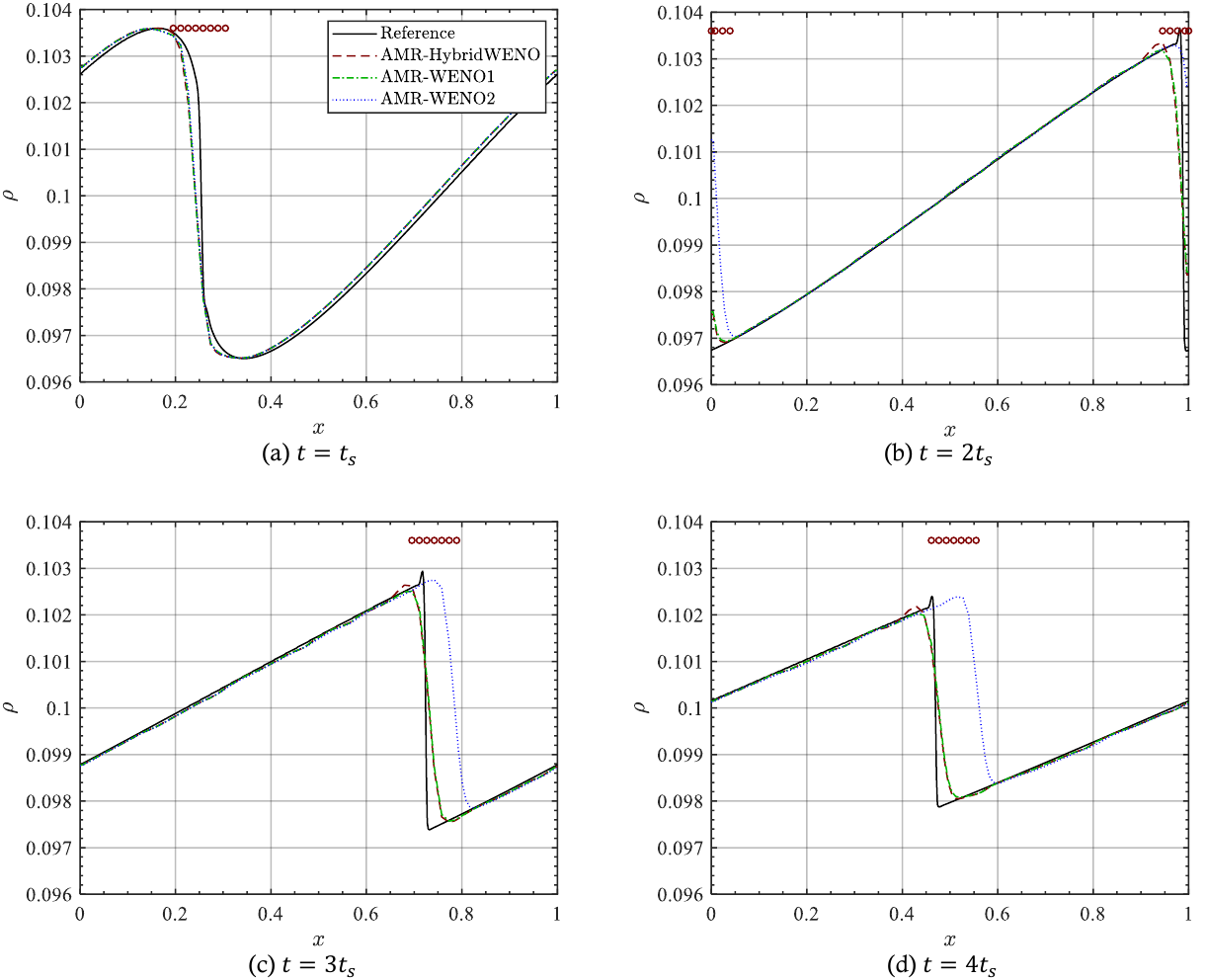}
    \caption{The acoustic wave problem at (a) $t = t_s$, (b) $t = 2t_s$, (c) $t = 3t_s$ and (d) $t = 4t_s$ on a 2-level fixed AMR grid. The reference solution is on a $N$ = 512 uniform grid. The AMR-HybridWENO, AMR-WENO1 and AMR-WENO2 solutions are obtained by hybrid interpolation, WENO interpolation with and without flux correction, respectively. The red circles indicates the troubled cell locations in AMR-HybridWENO method.}
    \label{fig:burgers}
\end{figure}

Among the compared interpolation approaches for problem a), Fig.~\ref{fig:burgers} illustrates that both WENO and hybrid WENO interpolations retain the correct shock locations at the sampled time instants during a long-period simulation, as long as the flux correction step is employed.
The hybrid WENO interpolation generates a result that is more aligned with the reference solution, with a small spike emerging behind the shock, whereas the WENO interpolation yields excessive nonlinear dissipation that damps this instability. 
The troubled cell technique in the hybrid interpolation strategy performs very well, by detecting the shock in sharp resolution irrelevant to its underlying grid levels.
In contrast, the WENO2 approach, i.e., WENO interpolation without flux correction, exhibits significant artifacts especially for long-period simulation, due to the non-conservation accumulation at the coarse-fine interfaces.
This scenario is very similar to the one in previous vertex-centered finite difference WENO studies in~\cite{sebastianMultidomainWENOFinite2003, tangNonconservativeAlgorithmsGrid1999}, where the nonconservative interface treatment causes flawed numerical shock locations.
The results here elucidates the superiority of the cell-centered framework over the vertex-based counterpart in the AMR calculations, where the nonconservative WENO interpolation can also give sufficiently high-accuracy results with the activation of flux correction at the coarse-fine interfaces.

\begin{figure}[htpb]
    \centering
    \includegraphics[width=0.9\linewidth]{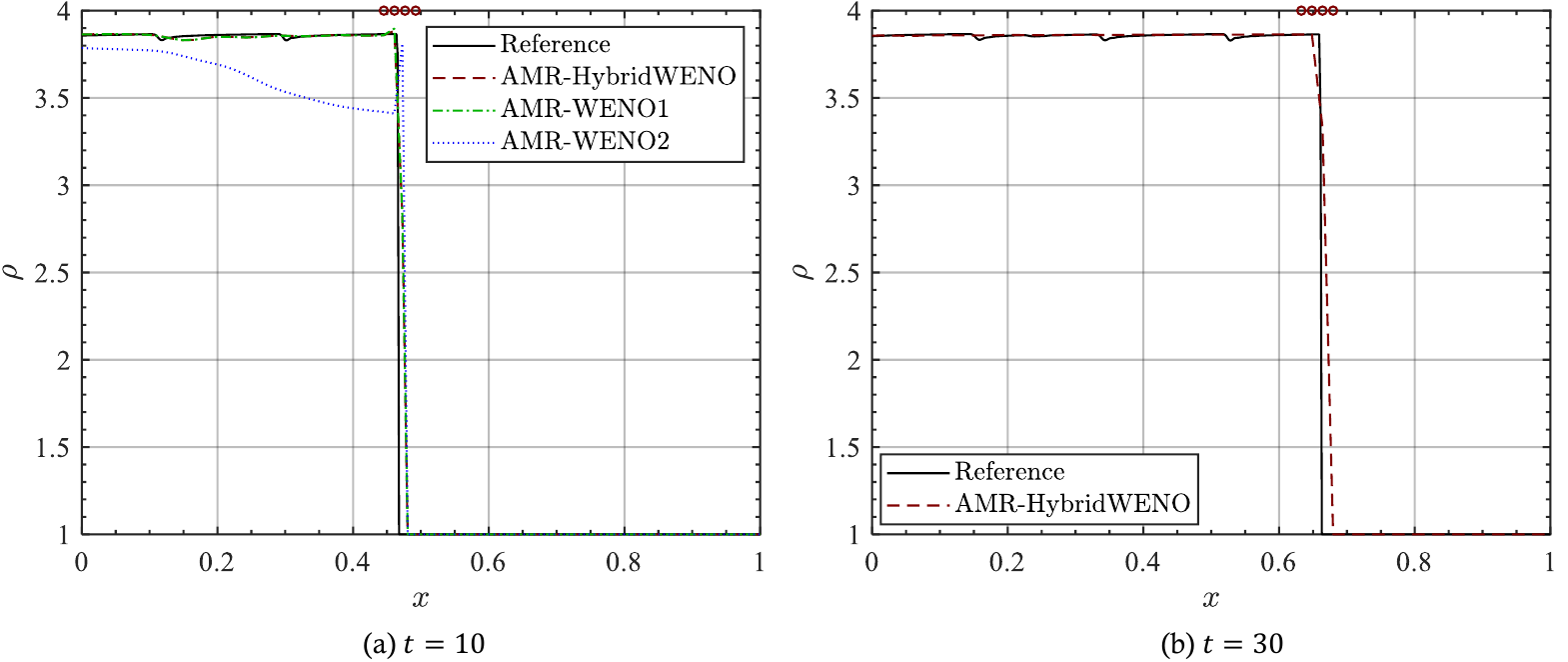}
    \caption{The slowly moving shock problem at a 2-level fixed AMR grid with a shock strength $Ma_s$ = 3 at (a) $t=10$ and (b) $t=30$. By the time $t=30$, the WENO interpolation methods terminate prematurely, so their results are not included in (b).}
    \label{fig:ma3-shock}
\end{figure}

The situation changes markedly in problem~(c), where a stronger shock interacts with the coarse–fine interfaces. 
As shown in Fig.~\ref{fig:ma3-shock}, the WENO2 interpolation begins to distort as soon as the shock passes through the first interface at approximately \( x = 4.7 \). 
Although the WENO1 approach is able to preserve the overall shock profile at time \( t = 10 \), noticeable oscillations appear in the post-shock region, indicating the onset of numerical instability. 
In the subsequent time steps, both WENO1 and WENO2 calculations fail and terminate prematurely. 
In contrast, the hybrid WENO interpolation remains stable throughout the long-time simulation and maintains sharp shock resolution at the final time \( t = 30 \).
For the pure WENO interpolation approaches, the failure is attributed to increased numerical instability in the presence of a strong shock. 
The non-conservative restriction procedure introduces erroneous values on the coarse cells, which are subsequently used as stencil points in the WENO reconstruction of neighboring coarse cells near the interface. 
This mechanism violates the Rankine–Hugoniot condition across the discontinuity and ultimately leads to numerical breakdown. 
In contrast, the hybrid interpolation strategy switches from the high-order non-conservative interpolation to a second-order conservative interpolation as soon as a shock is detected, thereby ensuring the stability of the computation.
It is also noteworthy that the periodic fluctuations observed in the post-shock region of the reference solution are significantly suppressed in the AMR solutions, which is likely due to the coarser grid resolution employed in the AMR configuration.

\subsection{One-dimensional Benchmarks}
\label{sec6.3}
The last section examined the shock–interface interaction problem with statically refined grids. 
In the following benchmarks, we consider more practical scenarios involving dynamically adaptive mesh refinement, 
where the refinement regions evolve with the time-dependent solution. 
It will further demonstrate the superiority of the proposed cell-centered AMR framework with the hybrid interpolation strategy over previous approaches.
\subsubsection{The Sod problem}
\label{sec6.3.1}
We consider the Sod problem with the initial condition
\begin{equation}
    (\rho, u, p)=\left\{\begin{array}{ll}(1,0,1), & x \in[0,0.5] \\(0.125,0,0.1), & x \in(0.5,1]\end{array}\right.
\end{equation}
Two sets of mesh are considered, by respectively a uniform grid as the reference solution and a 2-level AMR grid of a refinement ratio of 2.
They have the same grid resolution $\Delta x = 1/256$ on the finest grid.
Here, we identify a critical parameter $\textit{regrid\_int}$ which controls the time interval for performing a regrid during AMR that is used in AMReX.
For example, when $\textit{regrid\_int}$ = 1, the regrid procedure is performed at each step after the time marching on grid levels $\lambda$ with $\lambda<l$; for $\textit{regrid\_int}$ = 2, the regrid procedure is performed every two steps.

\begin{figure}[htpb]
    \centering
    \includegraphics[width=0.99\linewidth]{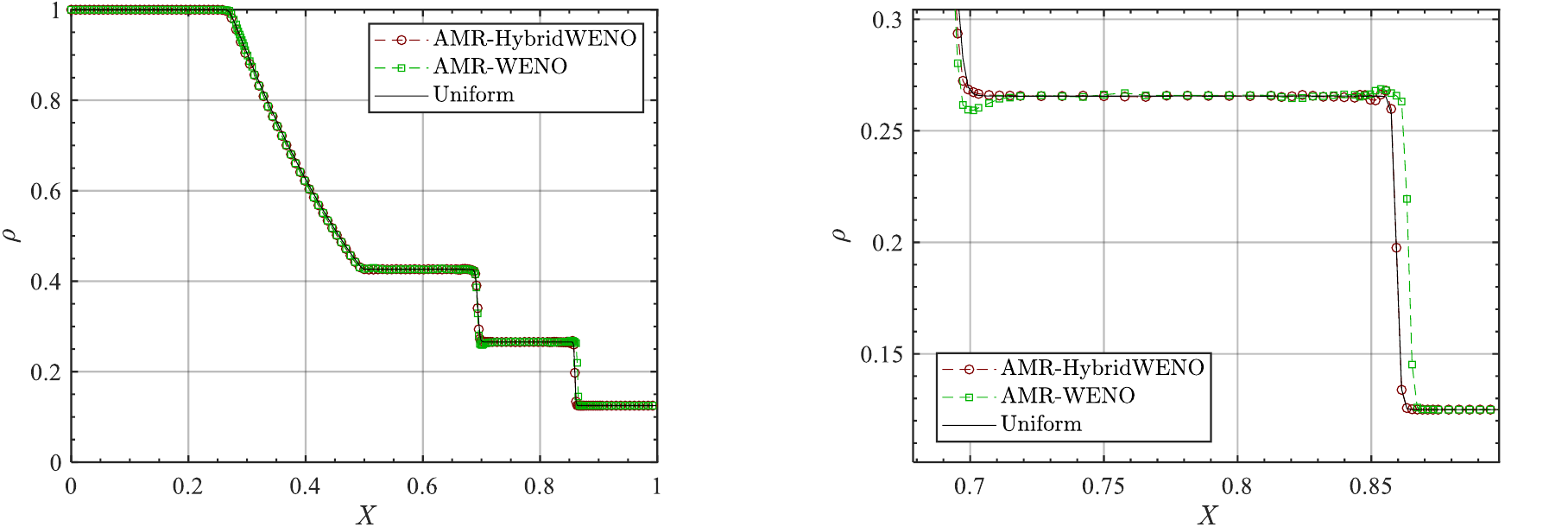}
    \caption{The Sod problem at $t$ = 0.2. For simplification, the troubled cell indication in hybrid WENO interpolation is omitted.}
    \label{fig:sod}
\end{figure}

At a $\textit{regrid\_int}$ of 2, we found that the results given by hybrid WENO and WENO interpolations exhibit negligible differences and both fit well with the reference solution, which is not presented here.
This finding agrees well with~\cite{shenAdaptiveMeshRefinement2011}, in which it is assumed that the discontinuity stays dynamically within the refined region.
While the non-conservative interpolation works well in this ideal scenario, as we increase $\textit{regrid\_int}$ to 5, the discrepancy occurs between the results by hybridWENO or WENO interpolation.
It is observed in Fig.~\ref{fig:sod} that the AMR-WENO method exhibits obvious oscillations at the front of shock and contact discontinuities, and most importantly, it produces a non-physical overshoot on the shock speed.
The AMR-HybridWENO method, on the other hand, produce very high-accuracy solution.

\subsubsection{The Lax problem}
\label{sec6.3.2}
\begin{figure}[htpb]
    \centering
    \includegraphics[width=0.99\linewidth]{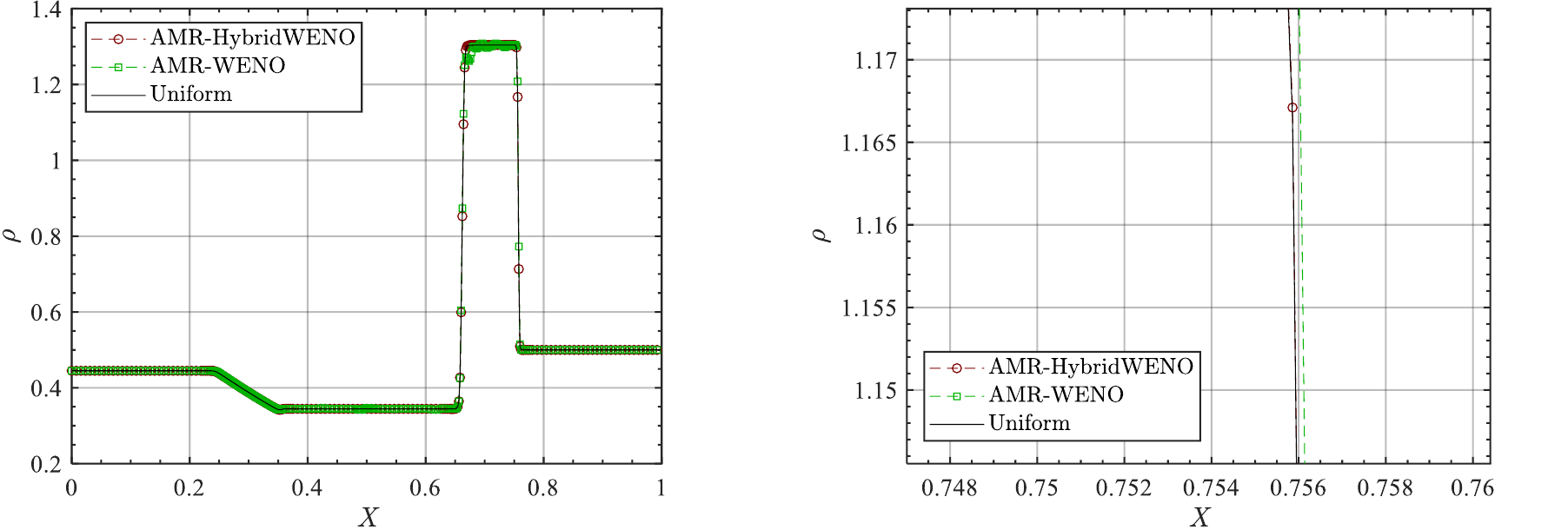}
    \caption{The Lax problem at $t$ = 0.1.}
    \label{fig:lax}
\end{figure}

We then consider the Lax problem with the initial condition
\begin{equation}
    (\rho, u, p)=\left\{\begin{array}{ll}(0.445,0.698,3.528) & x \in[0,0.5] \\(0.5,0,0.571) & x \in(0.5,1]\end{array}\right.
\end{equation}
In the case, we set the parameter $\textit{regrid\_int}$ = 10 to further test the robustness of interpolation schemes.
As illustrated in Fig.~\ref{fig:lax}, similarly, the non-conservative AMR-WENO method exhibits oscillations behind the material interface and produces a non-physical overshoot on the shock speed.
The AMR-HybridWENO method gives very stable and accurate results that agree well with reference results on a uniform grid.

% \subsubsection{The Sedov blast wave problem}
% This case illustrates the conservation errors with discontinuities, for both hybrid and nonconservative WENO interpolation methods.
% Maybe replace the former Lax problem case with this case.

\subsubsection{The Shu-Osher problem}
\label{sec6.3.3}
To illustrate the superiority of hybrid interpolation on smooth regions, we consider the Shu-Osher problem in which an initial shock interacts with the density sine waves.
The initial condition is
\begin{equation}
    (\rho, u, p)=\left\{\begin{array}{lr}(3.857143,2.629369,10.333333), & x \in[0,1] \\(1+0.2 \sin (5 x), 0,1), & x \in (1, 10]\end{array}\right.
\end{equation}

We compares different coarse-fine interpolation schemes by a) the original second-order one by Berger and Colella~\cite{bergerLocalAdaptiveMesh1989}, b) the HybridWENO interpolation presented in this work and c) the WENO interpolation.
The reference results are obtained on a uniform grid with $N$ = 2048.
It is observed that both the HybridWENO and WENO interpolations produce less dissipation in the smooth regions compared to the original second-order method.
This confirms the hybrid interpolation method presented in this work in switching between a high- and low-order interpolation schemes in this one-dimensional complex benchmark.

\begin{figure}[htpb]
    \centering
    \includegraphics[width=0.99\linewidth]{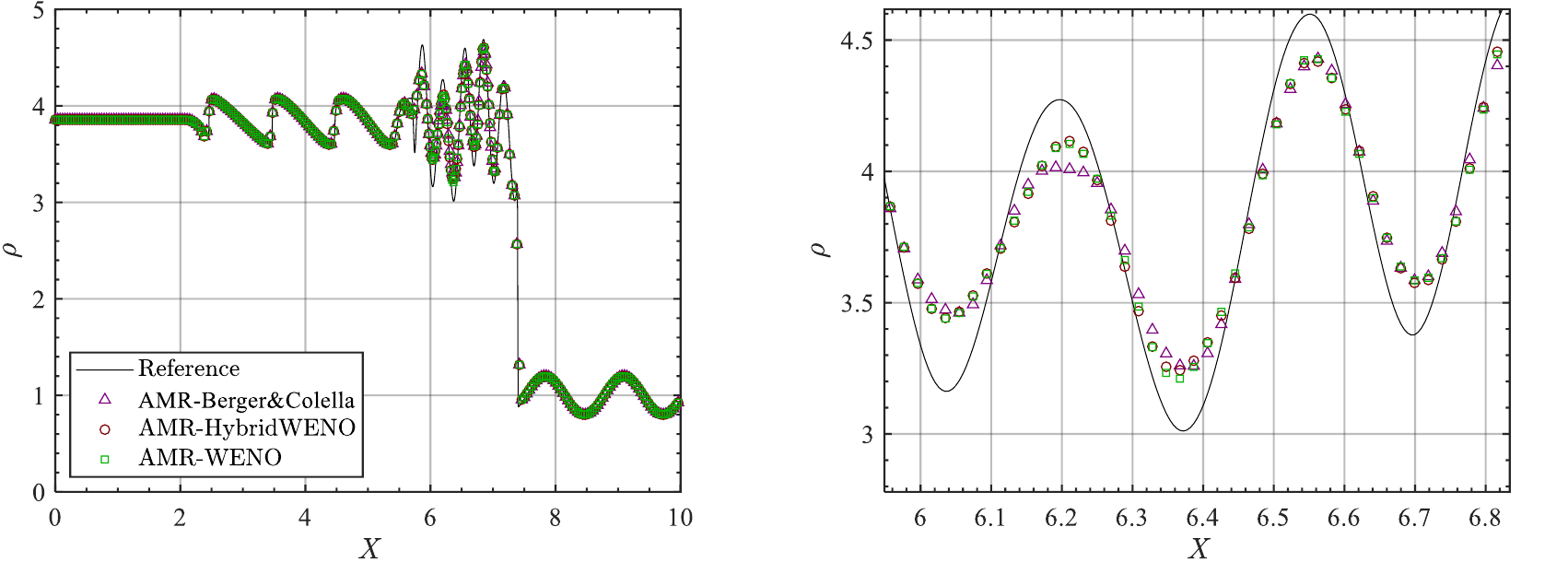}
    \caption{The Shu-Osher problem at $t$ = 0.2 on a 2-level AMR grid with a base grid of $N$ = 128. The reference result is obtained with a uniform grid of $N$ = 2048.}
    \label{fig:Shu-Osher}
\end{figure}

\subsection{Multi-dimensional Benchmarks}
\label{sec6.4}
% \subsubsection{Kelvin-Helmholtz instability}
% \label{sec5.3.1}
\subsubsection{The Riemann problem}
\label{sec6.4.1}
To test the method on handling complex multidimensional problems, we first consider a two-dimensional Riemann problem in a rectangular domain of $[0,1]\times[0,1]$, with the initial condition
\begin{equation}
    (\rho, u, v, p)=\left\{\begin{array}{ll}(1.5,0,0,1.5), & (x, y) \in[0.8,1] \times[0.8,1] \\(0.5323,1.206,0,0.3), & (x, y) \in[0,0.8) \times[0.8,1] \\(0.138,1.206,1.206,0.029), & (x, y) \in[0,0.8) \times[0,0.8) \\(0.5323,0,1.206,0.3), & (x, y) \in[0.8,1] \times[0,0.8)\end{array}\right.
\end{equation}

\begin{figure}[htpb]
    \centering
    \includegraphics[width=0.9\linewidth]{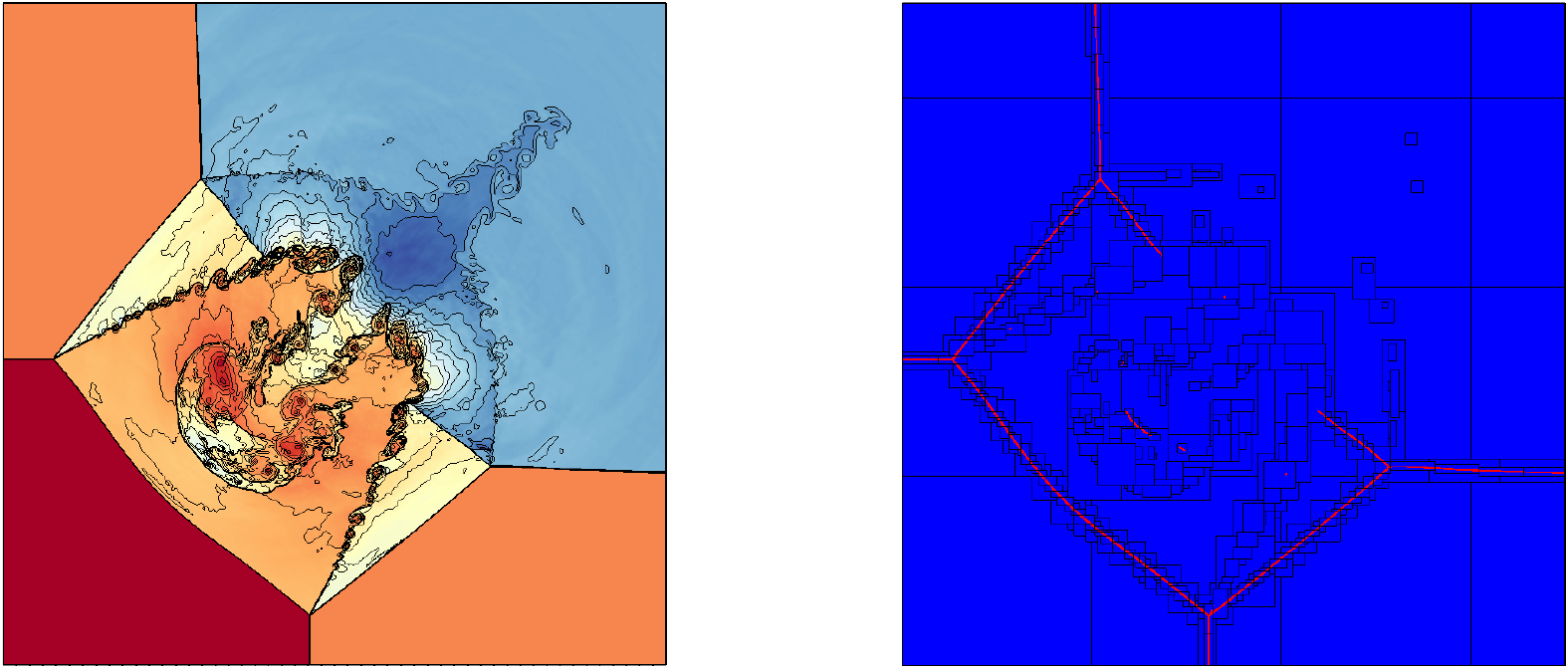}
    \caption{The two-dimensional Riemann problem at $t$ = 0.8 on a 3-level AMR grid with a base grid of $N_x\times N_y$ = 448 $\times$ 448. Left: the density, right: troubled-cell flagging and mesh refinement.}
    \label{fig:Riemann}
\end{figure}

The initial discontinuities are located along $x=0.8$ and $y=0.8$, and the simulation lasts until $t=0.8$.
A three-level AMR grid with the refinement ratio (2,2) is employed with the base grid $N_x\times N_y$ = 448 $\times$ 448, and the hybrid WENO interpolation scheme is used.
It is observed in Fig.~\ref{fig:Riemann} that the current method allows very sharp resolution on the subtle structures in the flow field.
The AMR technique works well by concentrating high-level grids on interested regions like the discontinuities and shear layers.
The troubled cell technique also performs well by identifying the shock structures with a considerable strength, including the local shocklet yielded by vortical turbulence, whereas the weak acoustic waves in the background are filtered through the Lax criterion.

\subsubsection{Double-Mach reflection}
\label{sec6.4.2}
We next consider the more challenging two-dimensional double-Mach reflection problem~\cite{woodwardNumericalSimulationTwodimensional1984}, 
in which an incident Mach~10 shock wave, with a specific heat ratio of \(\gamma = 1.4\), impinges on a reflecting wall at an angle of \(\theta = 60^\circ\). 
The pre- and post-shock states are specified as
\begin{equation}
\begin{aligned}
&[\rho_r,\ u_r,\ v_r,\ p_r] = [1.4,\ 0.0,\ 1.0,\ 1.0], \\
&[\rho_l,\ u_l,\ v_l,\ p_l] = [8.0,\ 8.25 \sin\theta,\ -8.25 \cos\theta,\ 116.5].
\end{aligned}
\end{equation}
The computational domain is \([0, 3] \times [0, 1]\). 
On the bottom boundary, a reflecting wall is imposed for \(x \geq 1/6\), while the region \(x \in [0, 1/6)\) is set to the exact post-shock state. 
The top boundary is prescribed to match the exact motion of the Mach~10 shock. 
The post-shock flow is imposed at the left boundary, and a zero-gradient Neumann condition is applied at the right boundary. 
To trigger the formation of the double-Mach reflection structure, the incident shock is initialized as a ramp located at \(x = 1/6\) with the profile 
\begin{equation}
x = \frac{1}{6} + \frac{y}{\tan(\pi/3)}.
\end{equation}

The base grid on the three-level AMR hierarchy consists of \(N_x \times N_y = 720 \times 240\) cells, with a refinement ratio of \((2,2)\). 
It is noted that both pure WENO and Lagrange interpolation fail immediately in this problem by producing negative density values. 
Similar observations were reported in~\cite{chenNonlinearWeightsShock2023,kamiyaApplicationCentralDifferencing2017}, where first-order interpolation was employed to maintain numerical robustness in the presence of strong shocks. 
In contrast, the hybrid WENO method exhibits better robustness until the final simulation time. 

\begin{figure}[htpb]
    \centering
    \includegraphics[width=0.6\linewidth]{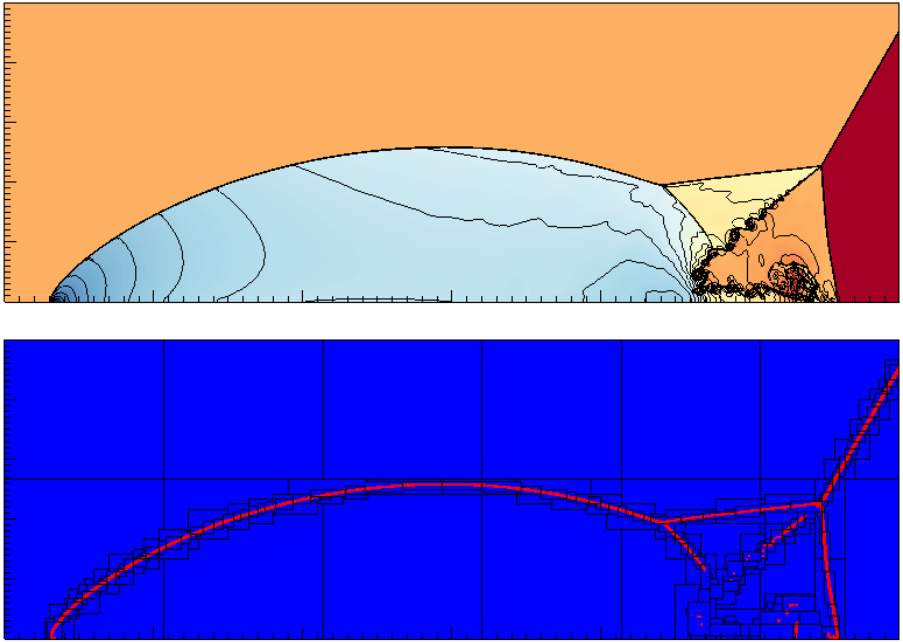}
    \caption{The double-Mach reflection problem at $t$ = 0.2 on a 3-level AMR grid with a base grid of $N_x\times N_y$ = 720 $\times$ 240.Top: density, bottom: troubled-cell flagging.}
    \label{fig:double-1}
\end{figure}

\begin{figure}[htpb]
    \centering
    \includegraphics[width=0.9\linewidth]{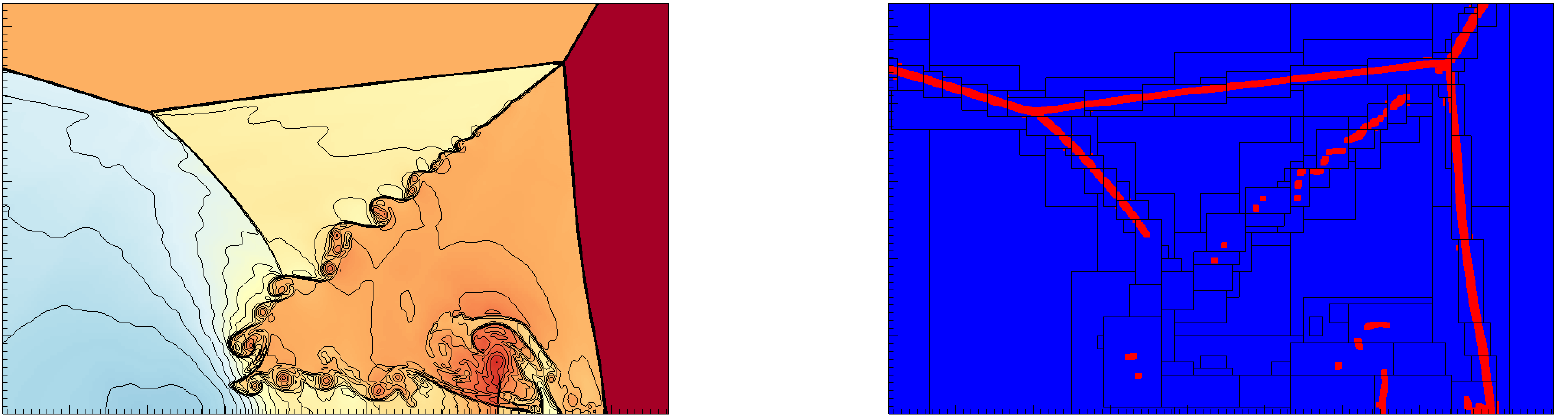}
    \caption{The double-Mach reflection problem at $t$ = 0.2 on a 3-level AMR grid with a base grid of $N_x\times N_y$ = 720 $\times$ 240, zoom-in pictures of Fig.~\ref{fig:double-1}.}
    \label{fig:double-2}
\end{figure}

As shown in Figs.~\ref{fig:double-1} and \ref{fig:double-2}, the unstable shear-layer structures are well captured by the AMR solution, and the troubled-cell detection accurately identifies the flow discontinuities including the eddy shocklets. 
The present method yields noticeably sharper shear layers and better-resolved vortex roll-up structures behind the reflected shock, as compared to~\cite{bergerLocalAdaptiveMesh1989}. 
This improvement is primarily attributed to the use of six-point central WENO stencils, which posses lower numerical dissipation.

\subsubsection{Decaying Isotropic turbulence}
\label{sec6.4.3}
The final test case considers decaying compressible isotropic turbulence with eddy shocklets~\cite{leeEddyShockletsDecaying1991}, whose locations are not known \textit{a priori}.
This benchmark is intentionally selected because the presence of intermittent shocklets produces the condition corresponding to scenario~\ref{item:3} as introduced in Section~\ref{sec5.1}, thereby providing a stringent evaluation of the deficiencies of current non-conservative interpolation approaches.
In this problem, the key dimensionless parameters characterizing the flow are the turbulent Mach number and the Taylor-scale Reynolds number
\begin{equation}
M_t = \frac{\sqrt{\langle u_i u_i \rangle}}{\langle c \rangle}, \qquad
Re_\lambda = \frac{\langle \rho \rangle u_{\mathrm{rms}} \lambda}{\langle \mu \rangle},
\end{equation}
where \(u_{\mathrm{rms}} = \sqrt{\langle u_i u_i \rangle / 3}\) is the root-mean-square velocity and \(\lambda\) is the Taylor microscale defined through 
\(\lambda^2 = u_{\mathrm{rms}}^2 / \langle (\partial u_1/\partial x_1)^2 \rangle\).
When the turbulent Mach number \( M_t \) is sufficiently high, weak shock waves, known as eddy shocklets, are generated spontaneously from the turbulent motions. 

The three-dimensional compressible Navier–Stokes equations are solved in a periodic domain \(\Omega = [0,2\pi]^3\) using a uniform grid with spacing \(\Delta x = 2\pi/64\) for preliminary validation, whereas the subsequent calculations are performed on a two-level AMR grid with a same base resolution. 
The physical viscosity follows a power law of the form
\begin{equation}
\mu = \mu_{\mathrm{ref}}\left(\frac{T}{T_{\mathrm{ref}}}\right)^{3/4},
\end{equation}
where the reference parameters are obtained from the initial condition.

The initial velocity field is generated as a random solenoidal field with an energy spectrum of the form
\begin{equation}
E(k) \sim k^4 \exp\big[-2(k/k_0)^2\big],
\end{equation}
where \(k_0 = 4\) is the most energetic wavenumber. 
The initial density, temperature and pressure fields are spatially uniform with
\begin{equation}
p_0 = 101325\,Pa, \quad T_0=300\,K, \quad \rho_0 = 1.1768\,kg/m^3.
\end{equation}

\begin{figure}[htpb]
    \centering
    \includegraphics[width=0.9\linewidth]{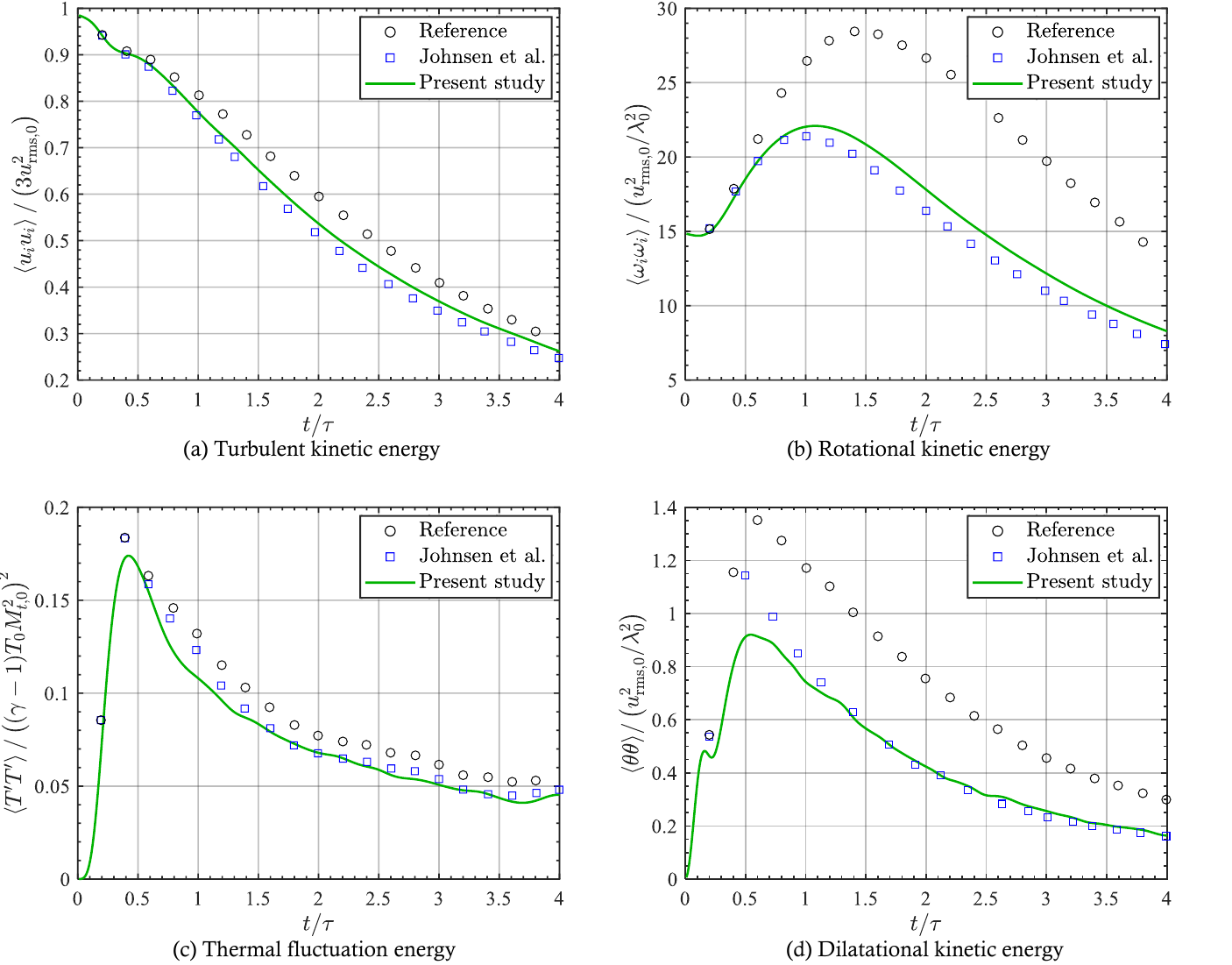}
    \caption{Temporal evolution of different energies for the isotropic turbulence problem on a $64^3$ uniform grid. The reference solution is calculated on a $256^3$ grid spectrally filtered to a $64^3$ grid, and the results from Johnsen et al. is obtained with the 7th-order WENO discretization for the inviscid terms~\cite{johnsenAssessmentHighresolutionMethods2010}.}
    \label{fig:HIT1}
\end{figure}

For preliminary validation of the numerical methods, the initial turbulent Mach number and Taylor-scale Reynolds number are set to \(M_{t,0} = 0.6\) and \(Re_{\lambda,0} = 100\), respectively, consistent with~\cite{johnsenAssessmentHighresolutionMethods2010}. 
The reference solution is obtained on a grid of size \(256^3\) and spectrally filtered to \(64^3\). 
The simulation is advanced to a nondimensional time of \(t/\tau = 4\), where \(\tau = k_0/u_{\mathrm{rms},0}\) is the initial eddy turnover time. 
The results are illustrated in Fig.~\ref{fig:HIT1}. 
For a fair comparison, the compared solution by Johnsen \textit{et al.}~\cite{johnsenAssessmentHighresolutionMethods2010} was obtained using a seventh-order WENO scheme for the inviscid terms and a sixth-order central difference scheme for the viscous terms, combined with a fourth-order Runge–Kutta time integration. 
It is observed that the present numerical methods produce results in good agreement with the reference solution, exhibiting comparable levels of numerical dissipation. 
A slight discrepancy is found in the dilatational kinetic energy, indicating that the present method is more dissipative in the compressible modes, which are particularly sensitive to numerical dissipation. 
Apart from this difference, other energetic quantities show excellent agreement. 
This serves as a preliminary validation of the proposed numerical methods for three-dimensional flows on uniform grids.

\begin{figure}[htpb]
    \centering
    \includegraphics[width=0.99\linewidth]{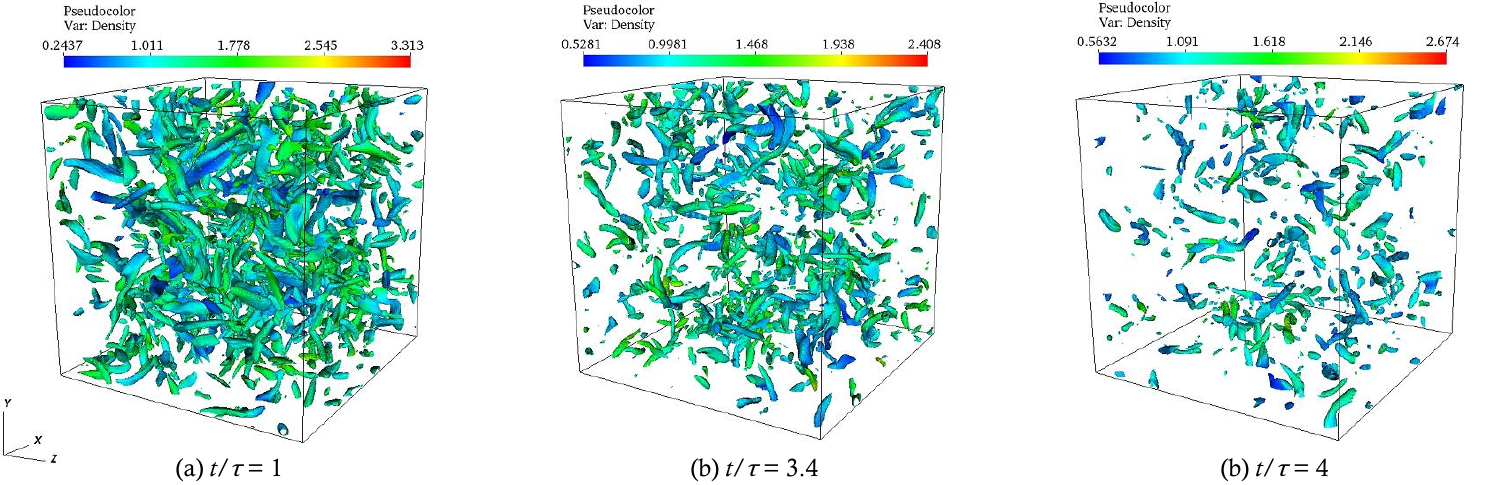}
    \caption{Three-dimensional turbulent coherent structures visualized by a Q-criteria iso-surface colored by the density at time (a) $t/\tau = 1$, (b) $t/\tau = 3.4$ and (c) $t/\tau = 4$, where $\tau$ is the eddy turn-over time.}
    \label{fig:HIT2}
\end{figure}

Next, we consider more challenging simulations on a two-level AMR grid. 
The refinement criterion is based on the dilatation field, with cells refined when \(\theta < \theta_t\), where \(\theta_t\) is a user-defined threshold. 
This criterion effectively identifies regions containing eddy shocklets, which are characterized by strongly negative dilatation values. 
Specifically, we simulate a case with an initial turbulent Mach number of \(M_{t,0} = 1.0\) to generate a more chaotic turbulent field populated with eddy shocklets. 
This setting is particularly useful for evaluating the robustness and resolution properties of different interpolation methods under conditions with stronger acoustic and entropy fluctuations. 
Figure~\ref{fig:HIT2} illustrates the instantaneous turbulent flow fields. 
The initially large eddies break down into smaller, randomly distributed vortices and gradually dissipate as time evolves.

\begin{figure}[htpb]
    \centering
    \includegraphics[width=0.9\linewidth]{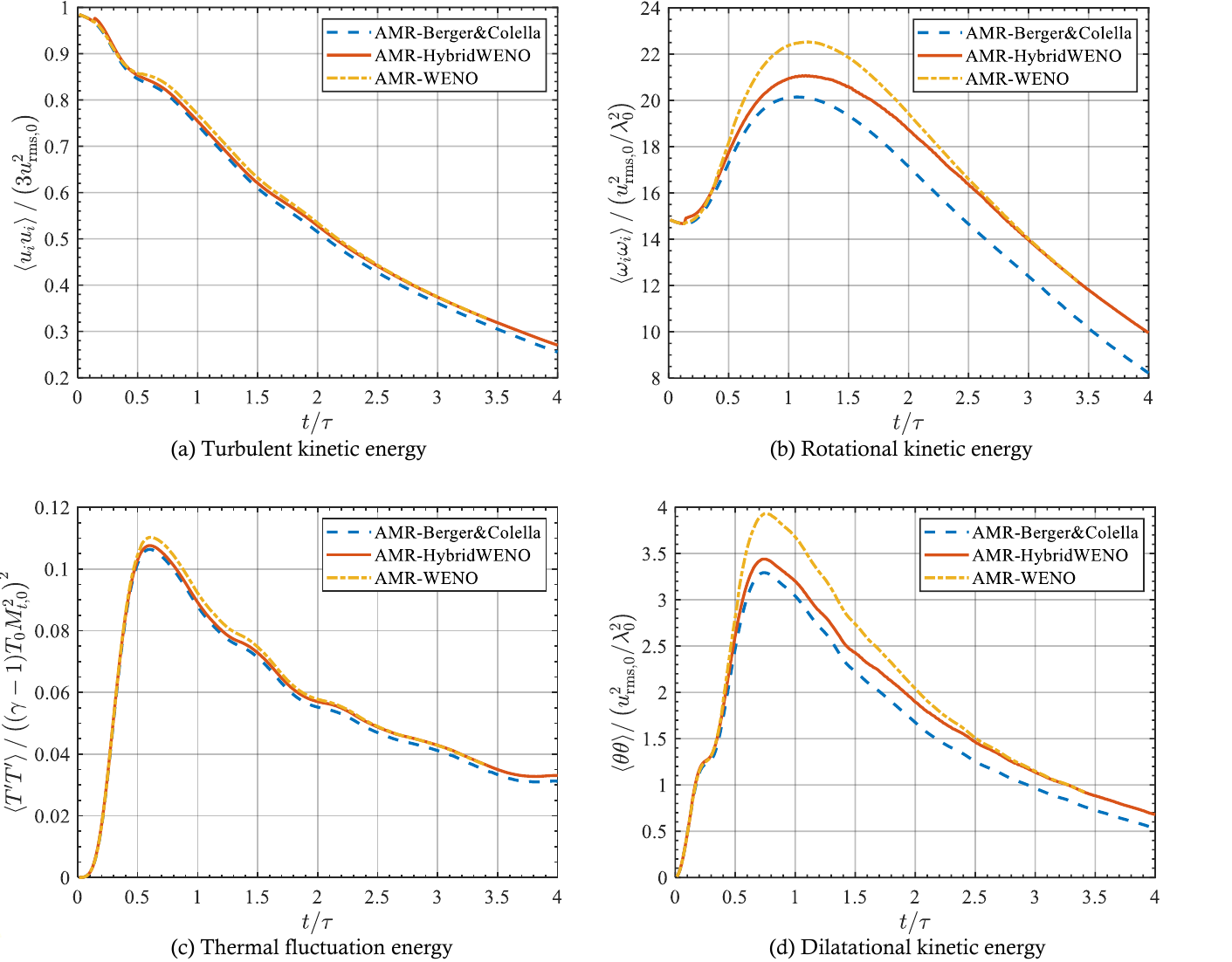}
    \caption{Comparisons among different interpolation strategies on the turbulence decaying on a 2-level AMR grid with a $64^3$ base grid.}
    \label{fig:HIT3}
\end{figure}

Figure~\ref{fig:HIT3} compares the evolution of energetic quantities during the decaying of isotropic turbulence for different interpolation methods. 
It is evident that the WENO interpolation retains the highest resolution among all methods, but suffers from severe numerical instability disappointingly. 
The computation using WENO interpolation terminates at approximately \( t = 3.4\tau \) due to this instability. 
In contrast, the low-order interpolation method exhibits the lowest resolution, as expected. 
The hybrid WENO interpolation lies between these two extreme.
On one hand, it achieves significantly higher resolution than the second-order interpolation.
on the other hand,  it also maintains great robustness throughout the entire simulation. 
As time progresses, its produced dissipation converges to that of the WENO interpolation, which is significantly better than the low-order interpolation method by Berger and Colella~\cite{bergerLocalAdaptiveMesh1989}.
This demonstrates its capability to achieve high resolution comparable to WENO while preserving numerical stability for long-term computations.

\begin{figure}[htpb]
    \centering
    \includegraphics[width=0.99\linewidth]{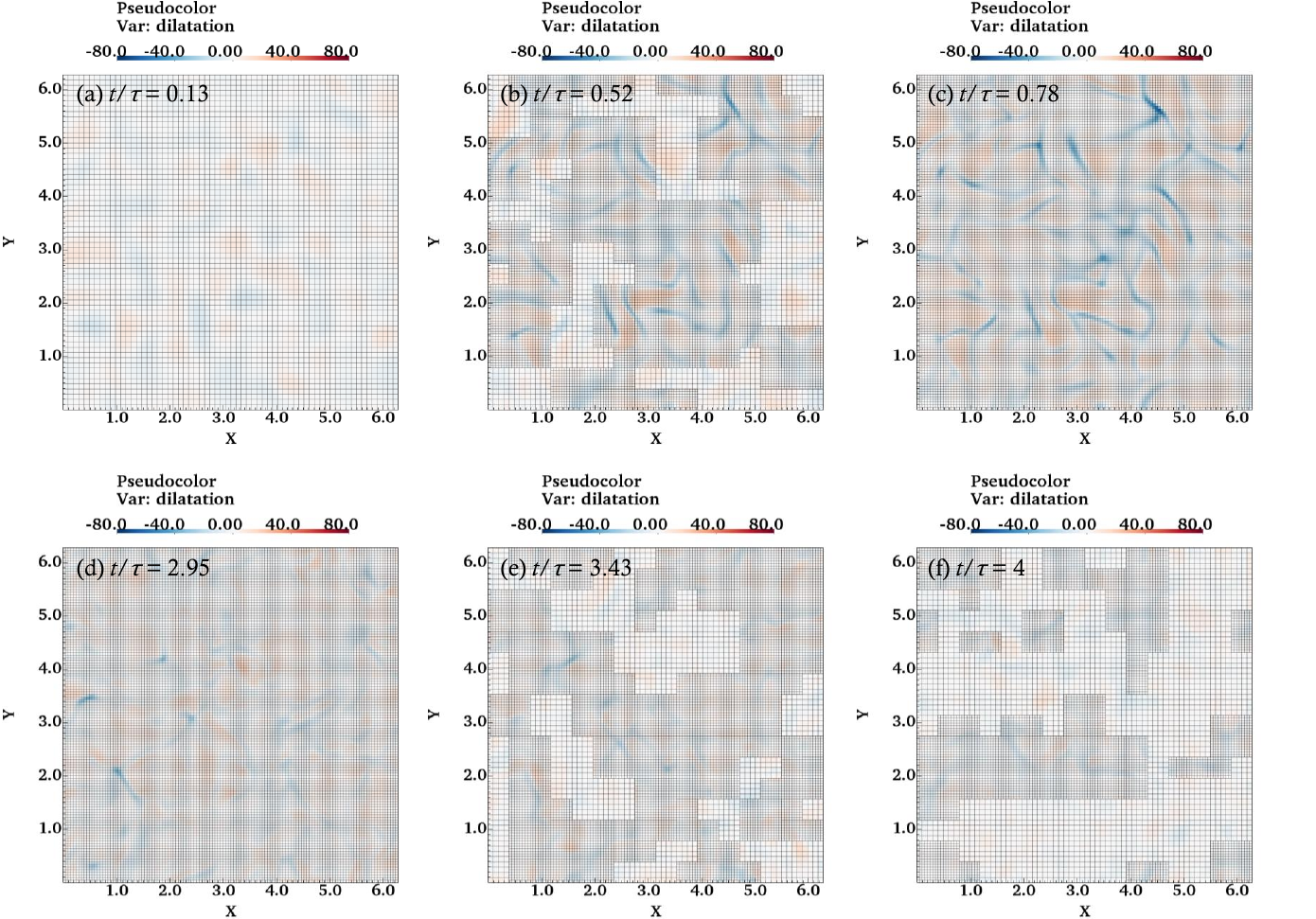}
    \caption{Temporal evolution of the eddy shocklets and the accompanied grid hierarchy distribution.}
    \label{fig:HIT4}
\end{figure}

To further investigate the failure mechanism of the WENO interpolation, Fig.~\ref{fig:HIT4} depicts a series of \(x\)–\(y\) plane slices of the dilatation field, superimposed with the adaptive mesh refinement. 
At the early stage of the simulation, the initially coarse grid is progressively refined as the smooth initial field rapidly develops into a highly turbulent state populated by randomly distributed eddy shocklets. 
By \( t/\tau = 0.78 \), the entire slice plane is refined, reflecting the widespread distribution of shocklets. 
Later on, as the turbulence decays, kinetic energy is converted into internal energy, leading to reduced compressibility and gradual dissipation of shocklets. 
Consequently, parts of the previously refined regions are coarsened, as illustrated in Figs.~\ref{fig:HIT4}(d)–(e). 
It is during this coarsening mechanism that the erroneous coarse solutions by non-conservative WENO interpolation are introduced into the computation, ultimately triggering the numerical instability. 
This mechanism is consistent with the abrupt termination for the AMR-WENO method as observed in Fig.~\ref{fig:HIT3}, where the WENO interpolation calculation fails at approximately the same time.

\section{Conclusions and limitations}
\label{sec7}
This paper addresses two principal issues for developing consistent high-order finite-difference AMR methods. 
The first is conservative errors at the coarse-fine interfaces, which is solved by adopting a cell-centered formulation, rather than the conventional vertex-centered counterpart in finite difference methods, for the construction of AMR solutions. 
The second is the introduction of a hybrid interpolation strategy, combined with a troubled-cell detection mechanism, to avoid non-conservative interpolation at the presence of discontinuities. 

New staggered-grid WENO interpolation schemes with reduced numerical dissipation are derived by employing a balanced central stencil for the prolongation and restriction steps, respectively. 
When used in conjunction with conservative interpolation/averaging, this interpolation strategy significantly facilitates the numerical accuracy and stability. 
Results from canonical benchmark problems demonstrate that the proposed method simultaneously preserves high-order accuracy in smooth regions and ensures local conservation across discontinuities. 
Moreover, the method exhibits strong robustness in several challenging problems that have posed difficulties for existing non-conservative AMR approaches. 
Since the framework is developed upon modern high-performance platform AMReX, it is naturally parallelized for both CPU and GPU architectures, making it well suited for large-scale simulations.

A current limitation of this work is the lack of a formal stability analysis for the hybrid WENO–conservative interpolation procedure. 
% Although the proposed approach demonstrates excellent robustness in practice, slight but tolerable oscillations are occasionally observed near the regions where the interpolation strategy switches between different modes. 
A more thorough theoretical investigation of this issue will be pursued in future work.

\newpage
\appendix
\section{High-order WENO-type restriction method for staggered AMR grids}
\label{app1}
For a given central stencil consisting of $2r$ uniformly spaced pointwise values, we define $(r+1)$ candidate substencils, each containing $r$ points. 
The $k$-th substencil is denoted as
\begin{equation}
S_k = \left\{ x_{i-r+k+s},x_{i-r+1+k+s},...,x_{i-1+k+s}\right\}, \quad k = 0, \dots, r-1.
\end{equation}
\begin{equation}
s = 0,1,...,r_i
\end{equation}
if we denote $r_i$ as the refinement ratio in this direction.

Once the point values on each substencil are determined, we construct the $(r-1)$-th order Lagrange interpolation polynomial on each substencil, which reads
\begin{equation}
u^{(k)}(x) = \sum_{m=0}^{r-1} c_m^{(k)}(x)\, u_{i-r+k+m+s},
\end{equation}
where the Lagrange basis functions are given explicitly by
\begin{equation}
c_{m}^{(k)}(x)=\prod_{\substack{m=0 \\ m \neq k}}^{r-1} \frac{x-x_{i-r+k+s}}{x_{i-r+k+s}-x_{i-r+k+s}}
\end{equation}

The global optimally $2r$-th order interpolant is then reconstructed as a convex combination of the substencil polynomials by
\begin{equation}
u_{\text{CWENO}}(x) = \sum_{k=0}^{r} \omega_k(x)\, u^{(k)}(x),
\end{equation}
with nonlinear weights $\omega_k(x)$ constructed to achieve both high accuracy and non-oscillatory behavior. Similar to the original WENO-JS scheme~\cite{jiangEfficientImplementationWeighted1996a}, we define
\begin{equation}
\omega_k = \frac{d_k}{(\epsilon + \beta_k)^p}, \qquad \tilde{\omega} = \frac{\omega_k}{\sum_{s=0}^{r-1} \omega_s},
\end{equation}
where $d_k$ are the optimal linear weights (exact for smooth data), $\beta_k$ are the smoothness indicators for each substencil, and typical choices are $\epsilon = 10^{-6}$ and $p = 2$.

The smoothness indicators are computed by integrating the squared derivatives of the substencil polynomial, with
\begin{equation}
\beta_k = \sum_{l=1}^{r-1} \int_{x_{m}}^{x_{m+1}} \left( \Delta x^{2l-1} \left( \frac{d^l}{dx^l} u^{(k)}(x) \right)^2 \right) dx,
\end{equation}
where $x \in [x_m, x_{m+1}]$ is the interpolation target region in AMR applications.

Unlike the fifth-order WENO interpolation used in Sebastian and Shu~\cite{sebastianMultidomainWENOFinite2003}, this central WENO method achieves optimally sixth-order accuracy with symmetry-preserving stencils. It provides improved spectral resolution and reduced dispersion in smooth flows, making it better suited for AMR applications where the accuracy across different AMR levels are critical.

Moreover, our formulation maintains still a minimum third-order accuracy even near discontinuities, providing consistent robustness and fidelity as the traditional fifth-order schemes. 
As a result, this method combines the best of both worlds: higher-order accuracy in smooth regions and non-oscillatory, stable behavior near discontinuities or complex AMR interfaces.

\bibliographystyle{elsarticle-num} 
\bibliography{ref.bib}




% %% For numbered reference style

%   \textit{\LaTeX: a document preparation system},
%   Addison Wesley, Massachusetts,

% \end{thebibliography}
\end{document}